\documentclass[journal]{IEEEtran}
\usepackage[OT1]{fontenc} 
\usepackage[noadjust]{cite}
\usepackage{siunitx}
\makeatletter
\def\input@path{{fig/}} 
\makeatother
\usepackage[pdftex]{graphicx}
\usepackage{amsmath}
\usepackage{comment}
\usepackage{booktabs}
\usepackage{array}
\usepackage{xcolor}
\usepackage{listings}
\usepackage[usestackEOL]{stackengine}
\usepackage[caption=false,font=footnotesize]{subfig}
\usepackage{url}
\usepackage{amssymb}

\DeclareMathOperator{\vecz}{Vec}
\newcommand{\subparagraph}{}
\usepackage{titlesec}
\usepackage{arydshln}

\titleformat{\paragraph}[runin]{\normalfont\bfseries}{}{0pt}{}[. \mbox{}]

\begin{document}
\title{MERIT: Tensor Transform for Memory-Efficient Vision Processing on Parallel Architectures}
\author{%
Yu-Sheng~Lin,~\emph{National~Taiwan~Univ.},
Wei-Chao~Chen,~\emph{Skywatch~Inc.}~and~\emph{Inventec~Inc.},
and~Shao-Yi~Chien,~\emph{National~Taiwan~Univ.}%
\thanks{Y.S. Lin and S.Y Chien are with the Graduate Institute of Electronics Engineering, National Taiwan University.}}
\makeatletter
\patchcmd{\@maketitle}
  {\addvspace{0.5\baselineskip}\egroup}
  {\addvspace{-1\baselineskip}\egroup}
  {}
  {}
\makeatother
\maketitle

\newcommand{\figrtext}{Fig.}
\newcommand{\tabrtext}{Table}
\newcommand{\lstrtext}{Listing}
\newcommand{\algrtext}{Procedure}
\newcommand{\secrtext}{Section}
\newcommand{\figref}[1]{\figrtext~\ref{fig:#1}}
\newcommand{\sfigref}[1]{\subref{fig:#1}}
\newcommand{\tabref}[1]{\tabrtext~\ref{tab:#1}}
\newcommand{\lstref}[1]{\lstrtext~\ref{lst:#1}}
\newcommand{\equref}[1]{(\ref{equ:#1})}
\newcommand{\algref}[1]{\algrtext~\ref{alg:#1}}
\newcommand{\secref}[1]{\secrtext~\ref{sec:#1}}
\newcommand{\ssecref}[2]{\ref{sec:#1:#2}}
\newcommand{\bA}{\mathbf{A}}
\newcommand{\bB}{\mathbf{B}}
\newcommand{\bC}{\mathbf{C}}
\newcommand{\bH}{\mathbf{H}}
\newcommand{\bI}{\mathbf{I}}
\newcommand{\bM}{\mathbf{M}}
\newcommand{\bR}{\mathbf{R}}
\newcommand{\bS}{\mathbf{S}}
\newcommand{\bT}{\mathbf{T}}
\newcommand{\bU}{\mathbf{U}}
\newcommand{\bX}{\mathbf{X}}
\newcommand{\bY}{\mathbf{Y}}
\newcommand{\ba}{\mathbf{a}}
\newcommand{\bb}{\mathbf{b}}
\newcommand{\bc}{\mathbf{c}}
\newcommand{\bd}{\mathbf{d}}
\newcommand{\bk}{\mathbf{k}}
\newcommand{\bm}{\mathbf{m}}
\newcommand{\bp}{\mathbf{p}}
\newcommand{\bt}{\mathbf{t}}
\newcommand{\bv}{\mathbf{v}}
\newcommand{\bw}{\mathbf{w}}
\newcommand{\bs}{\mathbf{s}}
\newcommand{\bx}{\mathbf{x}}
\newcommand{\bon}{\mathbf{1}}
\newcommand{\bzo}{\mathbf{0}}
\newcommand{\tsub}{\text{sub}}
\newcommand{\etal}{\textit{et al}. }
\newcommand{\ie}{\textit{i}.\textit{e}., }
\newcommand{\eg}{\textit{e}.\textit{g}. }
\newcommand{\Red}[1]{\textcolor{red}{#1}}
\newcommand{\stimes}{\mathsf{x}}

\begin{abstract}
Computationally intensive deep neural networks (DNNs) are well-suited to run on GPUs, but newly developed algorithms usually require the heavily optimized DNN routines to work efficiently, and this problem could be even more difficult for specialized DNN architectures.
In this paper, we propose a mathematical formulation which can be useful for transferring the algorithm optimization knowledge across computing platforms.
We discover that data movement and storage inside parallel processor architectures can be viewed as tensor transforms across memory hierarchies, making it possible to describe many memory optimization techniques mathematically.
Such transform, which we call Memory Efficient Ranged Inner-Product Tensor (MERIT) transform,
can be applied to not only DNN tasks but also many traditional machine learning and computer vision computations.
Moreover, the tensor transforms can be readily mapped to existing vector processor architectures.
In this paper, we demonstrate that many popular applications can be converted to a succinct MERIT notation on GPUs,
speeding up GPU kernels up to 20 times while using only half as many code tokens.
We also use the principle of the proposed transform to design a specialized hardware unit called MERIT-z processor.
This processor can be applied to a variety of DNN tasks as well as other computer vision tasks while providing comparable area and power efficiency to dedicated DNN ASICs.

\end{abstract}
\IEEEpeerreviewmaketitle

\section{Introduction}
\begin{figure*}[t!]
\centering
\subfloat[]{\includegraphics[scale=0.55]{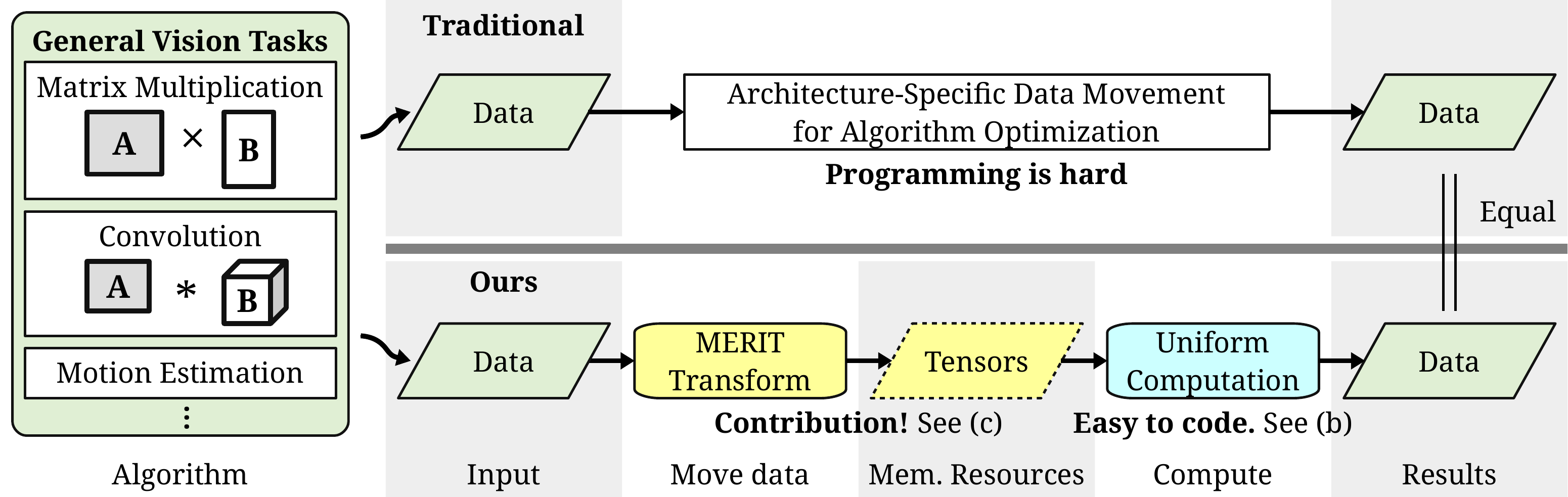}\label{fig:unr_as_fw1}}\\
\subfloat[]{\includegraphics[scale=0.45]{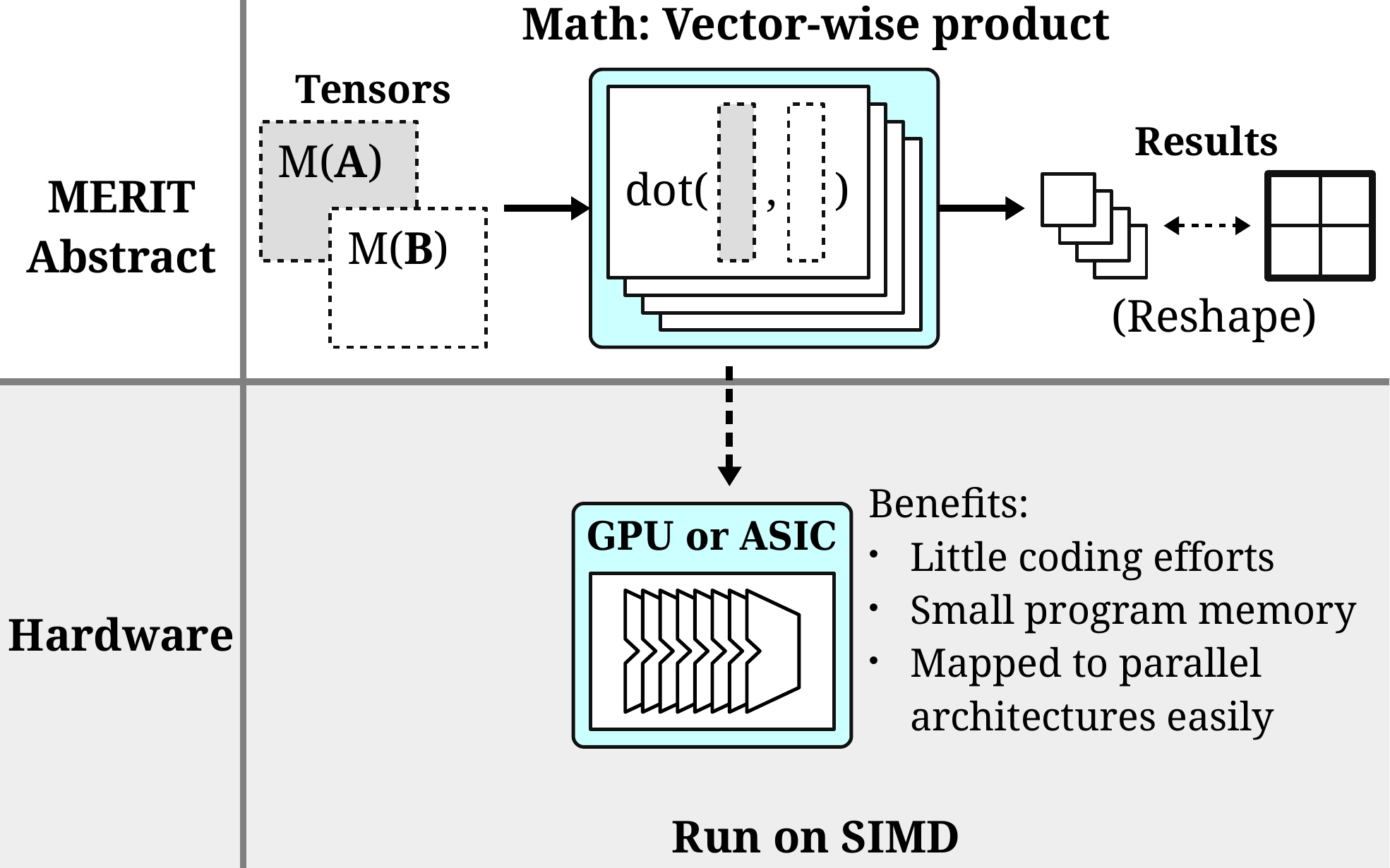}\label{fig:unr_as_fw2}}\quad
\subfloat[]{\includegraphics[scale=0.45]{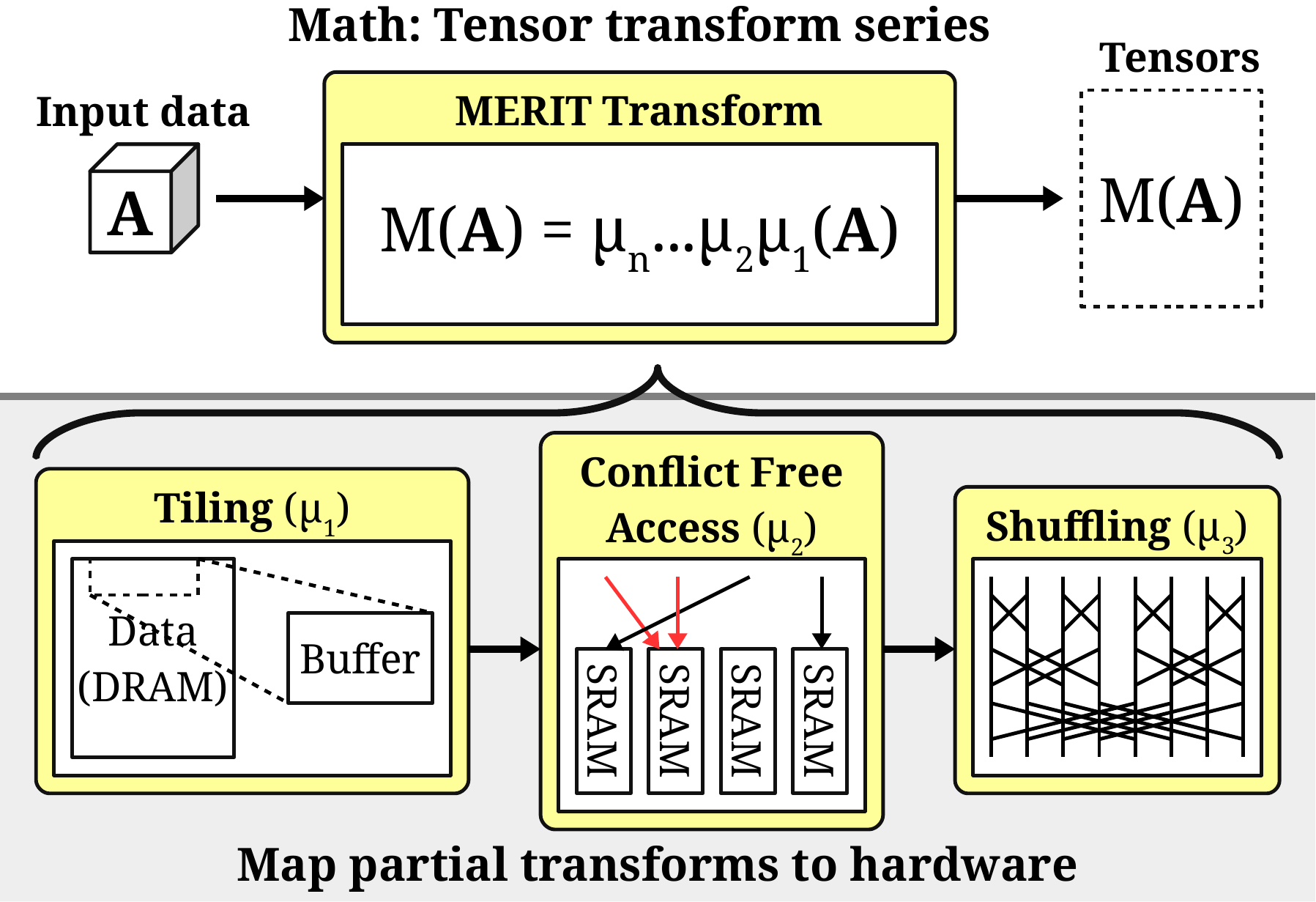}\label{fig:unr_as_fw3}}
\caption{\textbf{An overview of the MERIT transform.} (a) Reduce optimization efforts by abstracting the data movement stage into a tensor transform. (b) Convert computing tasks into a SIMD-friendly arrangement. (c) MERIT tensor can be mapped to hardware memory hierarchy for efficient execution.}\label{fig:unr_as_fw}
\end{figure*}

Recently, deep learning technology has gained remarkable success in many fields such as computer vision, image processing, and natural language processing.
The computations of deep neural networks (DNNs) tend to be fairly regular, and this means they are well-suited for highly-parallel processors such as general purpose graphic processing units (GPGPUs).
While GPU programming languages like CUDA and OpenCL have reached maturity and shown their promises in many scientific fields,
it still requires specialized knowledge in GPU architectures in order to implement DNN algorithms efficiently.
In some cases, libraries are implemented with insider knowledge that is not available to the public, such as cuDNN from NVIDIA~\cite{cudnn}.

For the purpose of efficient DNN computation, a recent trend is to create application-specific integrated circuits (ASICs), as they can often achieve lower power consumption or higher performance compared to general processors.
Popular designs include systolic arrays due to its ability to directly exchange data between processors~\cite{tpu,eyeriss} and 
GPU-like vector architectures with reduced arithmetic precision and larger on-chip memory~\cite{pudiannao}.
For extremely low power devices, ASICs can also help reduce memory pressure by processing compressed neural networks.
These techniques include binarization~\cite{xnornet,flexpoint}, dictionary-based compression~\cite{vq}, and pruning-based entropy compression~\cite{eie}.
While ASICs are more power efficient than GPUs, they are not flexible enough to keep up with the evolution of DNN techniques~\cite{selu,bnorm,maskrcnn,deconv,dilconv,deform,eva,flownet,mobilenet,pixel_shuf}.
The new types of computations can limit the widespread adoption of DNN ASICs since the arithmetic units and memory organization tend to be highly optimized for specific types of networks, which makes it difficult or impractical to map these new network layers to ASICs.

Even though the processors come in different varieties, the principles for optimizing data-intensive algorithms are always the same -- bring data closer to the processing elements.
For this purpose, DNN processors often do not employ traditional memory hierarchies.  Instead, programmers are given the freedom to explicitly control data movement between different types of memories, such as local, on-chip, or off-chip memory buffers.
Unfortunately, this means each algorithm needs to be optimized for each new processor by redesigning a new data movement procedure, causing repeated and wasted efforts.

To solve this problem, we propose the Memory Efficient Ranged Inner-Product Tensor (MERIT) transform which can incorporate data movement patterns into algorithms, as shown in \figref{unr_as_fw1}.
A MERIT transform converts one matrix (or generally, one tensor) into another by performing permutation and repetition of the matrix elements.
This allows us to convert computational tasks into vector-wise products or reductions, as shown in \figref{unr_as_fw2}.
An efficient MERIT transform therefore conforms to the steps of data journey through the memory hierarchy, and each step perform a partial transform representing a data movement procedure, such as tiling, shuffling, or bank conflict avoidance, as shown in \figref{unr_as_fw3}.
These partial transforms correspond to common resource constraints of GPUs or DNN architectures, such as limited on-chip SRAM or partially connected data networks.
Therefore, when a general vision-related computation can be reduced to a MERIT transform, many memory optimization techniques can be implemented on a particular architecture with very little effort.

To demonstrate the usefulness of MERIT, we use it to optimize several algorithms on GPUs and ASICs,
including bilateral filter~\cite{bilateral}, motion estimation, and many DNN layers~\cite{alexnet,vgg,saliency,pixel_shuf,dilconv,flownet,mobilenet}.
For GPUs, we implement MERIT for CUDA as a header file,
which can be easily integrated into existing algorithms to provide higher performance than many hand-tuned implementations.
The simplicity of the MERIT representation, and its strong connection between mathematics and architecture, means that we can also use it to create hardware with relatively little cost and effort.
We adopt this approach to design an open-source ASIC processor, the MERIT-z, that uses classic, efficient circuits such as butterfly networks to achieve efficient DNN and computer vision performance.

Specifically, our contributions are as follows:
\begin{itemize}
\item MERIT transform -- a data transform methodology that can succinctly describe data movement and reduce optimization efforts in parallel processing,
\item MERIT for CUDA -- an open-source API which produces fast GPU kernels with fewer code tokens compared to n\"aive CPU implementations, and
\item MERIT-z processor -- an open-source general vector processor designed with insights gained from the MERIT transform, supporting both common DNN layers and traditional vision processing.
\end{itemize}

In \secref{rel}, we shall next build up the background knowledge for readers.
In \secref{pro}, we formally define the MERIT transform and illustrate how to express different vision workloads with the transform.
Based on such transform definition, in \secref{imp} and \ref{sec:proc}, we show how to map parallel workloads to an efficient CUDA execution and onto the MERIT-z processor.
Finally, we show the experiments results and sum up the paper.

\section{Related Works}\label{sec:rel}
\paragraph{Kernel transform for parallel program optimization}
For parallel architectures, the performance of a computation task is usually limited by the memory bandwidth to the computing units.
While there are many common principles for memory optimization,
it has never been easy to optimize an algorithm on a particular parallel architecture, and there has been extensive research to tackle this problem.

One approach is to design a new, \textit{domain-specified programming language} (DSL) that can describe applications in specific domains, such as Halide~\cite{halide13} and TVM~\cite{tvm}.
An algorithm written in DSLs usually requires fewer lines of code than those in generic languages like C++ or OpenCL.
DSLs can also be compiled to target specific types of hardware,
and the process often involves searching for parameters for different memory configurations like~\cite{matog}.
This means the design of DSL compilers and their searching strategies require significant efforts,
and their language features are less stable because they can evolve more rapidly compared to generic languages.
Some DSLs also involve advanced programming concepts, making it difficult appeal to the wider audience.
For example, Halide defines images as N-dimension functions and describes the image pipelines as functional compositions.
In TVM, the computation schedule is stored in symbolic variables for programmers could use to build algorithms.

Another simpler approach is to convert a new computation problem into an existing one.
For example, a common implementation for the CONV layer in deep learning on GPUs is to unroll, or flatten,
patches of an image to columns of a matrix.
This transforms the CONV layer into a GEneral Matrix-Matrix multiplication (GEMM),
a routine that has been heavily optimized in existing libraries such as cuBLAS.
This technique has also been applied to ASIC design in order to support CONV on GEMM architectures~\cite{diannao,tpu}.
Conversion-based methods are easier to implement
since programmers only need to write the conversion routine to take advantage of the well-optimized routines.
However, not every problem can be easily converted,
and the process may sometimes introduce significant overhead which shall be discussed further in \secref{pro}.

The third approach is to divide algorithms into combinations of interchangeable function blocks.
Programmers implement algorithms by connecting different blocks without having to understand how to actually schedule the computation.
For example, in a GPU,
programmers write the \textit{vertex} and \textit{fragment shaders} to describe object transform and color shading rules,
and the 3D objects are rendered to the 2D screen automatically by the processor.
MapReduce~\cite{mapreduce} decomposes large-scale text-processing tasks into data sorting and two custom functions, \textit{map} and \textit{reduce}.
Convolution Engine~\cite{cengine} is a MapReduce-based streaming image processor with a restricted map and reduce operators.
UMI Operator~\cite{umi} shows several vision-related computations can be described by a custom unrolling procedure and a generalized inner-product function.

Later in this paper, we show that, for many vision related kernels or algorithms, 
we can decompose their data movement processes such that algorithms can run on simple SIMD architectures.
This transform is related to the unrolling above and can be defined with a few parameters for the data network hardware to execute efficiently.
We also propose a methodology for mapping this transform to GPUs and ASICs, which can greatly reduce the optimization effort for parallel programs.

\paragraph{Hardware for image, vision, and DNN}
Here we review several DNN, machine learning accelerators, and vision processors published in mainstream conferences for graphics, computer architecture, and circuit technology.
For a more comprehensive review of DNN accelerators, we recommend the readers to consult survey papers such as \cite{nn_list, nn_list_git}.

Dedicated DNN ASICs are usually designed for speeding up either FC or CONV layers,
but they can also support other types of layers via conversion.
For example, DianNao~\cite{diannao} uses a ($16\times 16$)-by-($16$) matrix-vector processor to compute both FC and CONV layers, and it needs to unroll the feature maps in the input buffers in order to evaluate the CONV layers.
TPU~\cite{tpu} uses systolic arrays and is able to perform a variable ($256\times 256$)-by-($256\times N$) matrix-matrix operation, and it supports CONV layers by storing the unrolled feature maps within $256$ separate queues.
Eyeriss~\cite{eyeriss} holds a row in the processor and adds a path to broadcast rows for the CONV layers, but this path is not utilized by the FC layers.
In a more recent work~\cite{DNPU}, separate hardware units are used to process FC and CONV layers independently.
MAERI~\cite{maeri} adopts a tree-based distribution and reduction path to support more types of DNN layers,
but it can be difficult to configure these trees.

There exist ASICs designs that aim to support more general parallel computation tasks.
Convolution Engine~\cite{cengine} and PuDianNao~\cite{pudiannao} use the fixed function pipeline architecture and support different types of computations by enabling different functional units.
Reconfigurable interconnection is yet another example, where 
FPGAs are dynamically programmed to execute various computing tasks~\cite{darkroom,catapult}.
In CRISP~\cite{CRISP} streaming processor, different functional units are connected by a programmable crossbar,
and various image processing methods could be applied as image pixels are streamed into the processor.

Register File Cache prevents the round trip between processors and the SRAM, improving the overall throughput and reducing the power consumption~\cite{rfc}.
Affine Warp detects the regular register values across parallel threads, minimizing the redundancy of computing and storing these registers~\cite{affinecomp}.
The design of our MERIT-z processor is related to these GPGPU optimization research, except that our architecture is derived through insights from the mathematical framework presented in this paper.

\section{The MERIT Transform}\label{sec:pro}
%
%
%

\begin{figure}[t]
\centering
\includegraphics[width=0.49\textwidth,page=3]{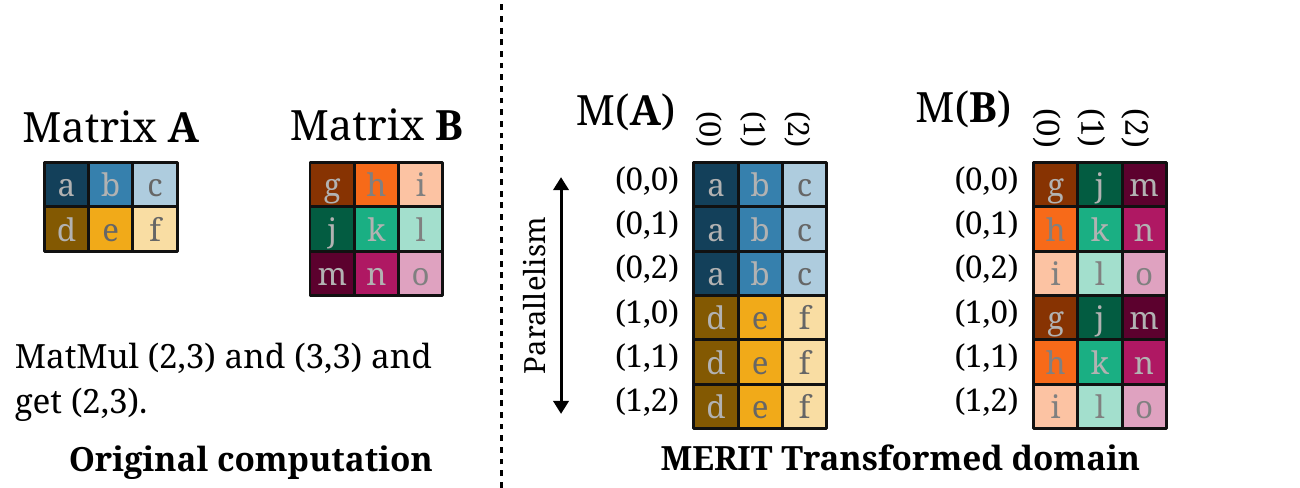}
\caption{A matrix multiplication operation in the parallel form using MERIT.}\label{fig:merit_unrm}
\end{figure}

The MERIT transform, denoted as $M(\cdot)$, is a mathematical process for converting a tensor into another,
such that a given algorithm can be transformed into the SIMD domain for easier operation.

As an example, we can apply the MERIT transform to the GEMM algorithm $\bC = \bA\bB$ and obtain

\begin{equation}\label{equ:umi}
\vecz (\bC) = R(M(\bA), M(\bB), \odot),
\end{equation}
where $R(\bX, \bY, \odot)$ means applying dot-product $\odot$ to every row of $(\bX, \bY)$, $\vecz(\bC)$ is the vectorized form of matrix $\bC$, 
and the MERIT transforms $M(\bA)$, $M(\bB)$ are created by repeating rows or columns of the input matrices (\figref{merit_unrm}).
While this formulation appears to complicate matters, there are several computational advantages to this expression.
First, each dot-product can be computed independently, which means this equation can be evaluated in parallel.
Second, $M(\cdot)$ and $\odot$ can both be specified by the users, making this equation configurable for general computational tasks.
Third, elements in the transformed matrices are copies of the original ones, which means $M(\cdot)$ is a pure data movement operation, leaving $\odot$ as the only real arithmetic operations in the equation.

Because the MERIT transform $M(\cdot)$ implies data movement, 
it is important to make sure $M(\cdot)$ maps to actual hardware to achieve optimal performance.
For this purpose, we factorize the MERIT transform $M(\bA) = \mu_n \cdots \circ \mu_2 \circ \mu_1(\bA)$,
where each partial transform, or sub-step $\mu_i$, is a repetition and permutation on the input data (\figref{unr_as_fw3}).
For a particular architecture, we would need a particular set of sub-steps to reflect its memory and data-path design such as multi-bank SRAM or DRAM.
Fortunately, these sub-steps can often be reused for the same architecture across different algorithms, and 
we shall discuss the subject of efficient MERIT transform in \secref{imp}.
Before then, let us start with a few more examples to familiarize the readers with the benefits of this transform.

\subsection{Convolution on SIMD Processors}

\begin{figure}[t]
\centering
\subfloat[Convolution via unrolling.]{\includegraphics[width=0.49\textwidth,page=1]{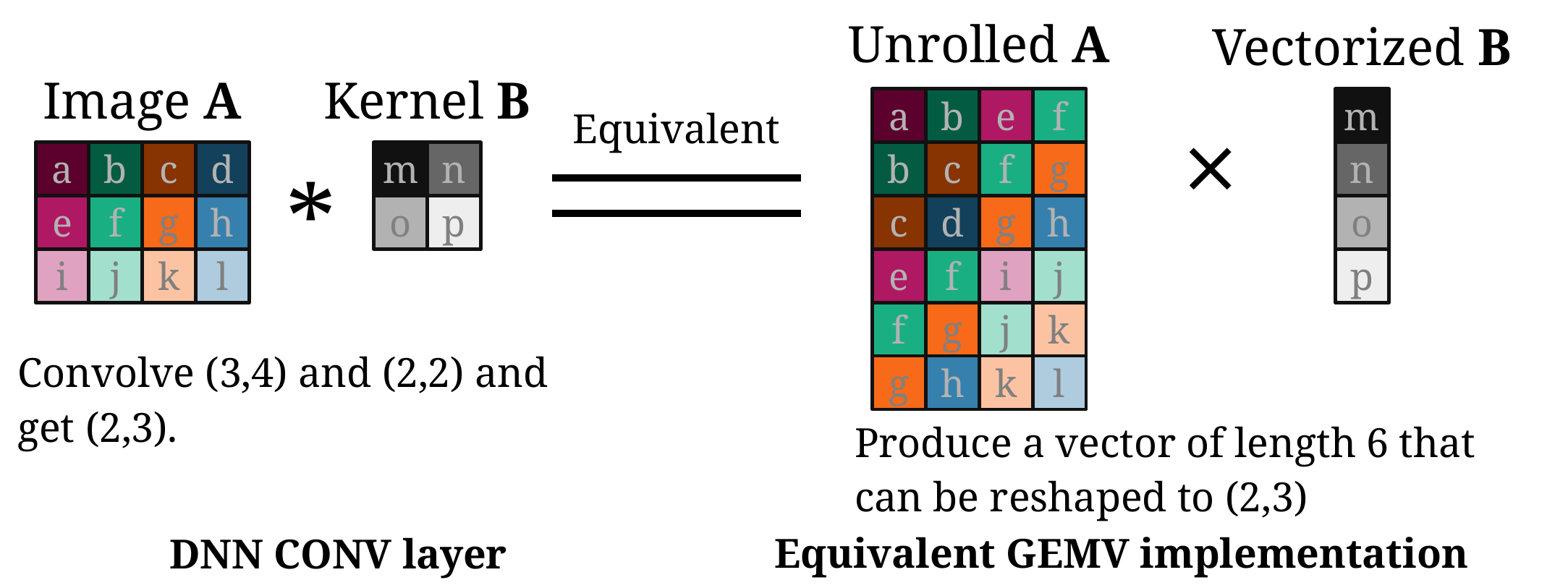}\label{fig:naive_unrc}}\\
\subfloat[Convolution in MERIT representation.]{\includegraphics[width=0.49\textwidth,page=2]{UMI_gemmconv.pdf}\label{fig:merit_unrc}}
\caption{A convolution operation in parallel form.}\label{fig:compare_conv}
\end{figure}

\figref{naive_unrc} shows a popular technique for accelerating convolution on SIMD processors~\cite{caffe}.
This process converts convolution into GEneral Matrix-Vector multiplications (GEMVs) by unrolling the input image $\bA$ and 
vectorizing the kernel $\bB$, which can be written as
\begin{equation}\label{equ:naive_unr}
\vecz (\bC) = U(\bA) \times \vecz (\bB).
\end{equation}
This is very effective because it takes advantage of the well-optimized GEMV routines on SIMD processors.
However, unrolling creates an unnecessarily large input matrix $U(\bA)$, which is clearly sub-optimal.

A more optimal process would be to replicate data on-the-fly.
This means we could assign $M(\bA)\equiv U(\bA)$, and expand the data during the transfer process, as shown in \figref{unr_as_fw3}.
Rather than expanding the image before the input stage, $M(\cdot)$ expands during data movement, such that replication of data can occur as late as possible to reduce memory bandwidth and buffer size.
Note that in order to map the above equation to \equref{umi}, $M(\cdot)$ also needs to be applied to the kernel $\bB$ (\figref{merit_unrc}), but this is efficient and also done on-the-fly.


\subsection{Extending to General Tensor Product}

A MERIT transform is coupled with a product function $\odot$ such that \equref{umi} represents an algorithm and can run on SIMD efficiently.
We can change this function to adapt the equation to different algorithms.
For the MERIT GEMM and convolution, we apply the dot-product
\begin{equation}\label{equ:relu_vprod}
\bv_1 \odot \bv_2 \equiv \bv_1^T\bv_2
\end{equation}
to each vector pair of the transformed matrices. Similarly, if we define
\begin{equation}
\bv_1 \odot \bv_2 \equiv
\left\lbrace
\begin{aligned}
& \max(\bv_1^T\bv_2, 0)\\
& ||\bv_1 -\bv_2||_1
\end{aligned}
\right. ,
\end{equation}
then we apply our transform to more computations,
such as fused ReLU layer or patch distance algorithms like motion estimation.
If we couple such vector products to MERIT transform,
then we in fact apply vector functions to flattened tensors from 2D image patches or 3D convolution kernels.

Instead of defining the function $\odot$ on vectorized tensors $\vecz(\bT_1)\odot\vecz(\bT_2)$,
a more natural way is to define a tensor product that computes $\bT_1\odot\bT_2$ directly.
Our proposal for a generalized tensor product, called \textit{Ranged Inner-Product}, is related to the ideas from~\cite{umi,cengine} where
a vector product is expressed as a \textit{strategy class}, one of the most common \textit{design patterns}.
An implementation of the strategy-based vector product is listed below in \lstref{inner_st}. 

\begin{lstlisting}[language=c++, caption={$\odot$ for the CONV+ReLU layer, as a strategy class.}, label=lst:inner_st]
class ReLUStrategy {
   float act;
   void PreLoop() {act = 0;}
   void Loop(float a, float b) {act += a*b;}
   float PostLoop() {return max(act,0);}
};
typedef vector<float> Vector;
template<class Strategy>
float inner_prod(Vector a, Vector b) {
  Strategy s;
  s.PreLoop();
  for (int i = 0; i < a.size(); ++i)
    s.Loop(a[i], b[i]);
  return s.PostLoop();
}
// This line actually calls the odot function
inner_prod<ReLUStrategy>(...);
\end{lstlisting}

Here, programmers can write the ReLU product as a \textit{class} consisting of three \textit{strategy functions} and a \textit{for-loop}.
These strategy functions perform partial sum initialization, accumulation, and clipping, respectively.
Such strategy class, when combined with MERIT transform,
provides high flexibility for expressing various workloads.
For example, if we load another tensor and add it to \texttt{act} in \texttt{PostLoop},
we then obtain an inner-product function for residual layer.

\begin{figure}[t!]
\centering
\includegraphics[width=0.49\textwidth]{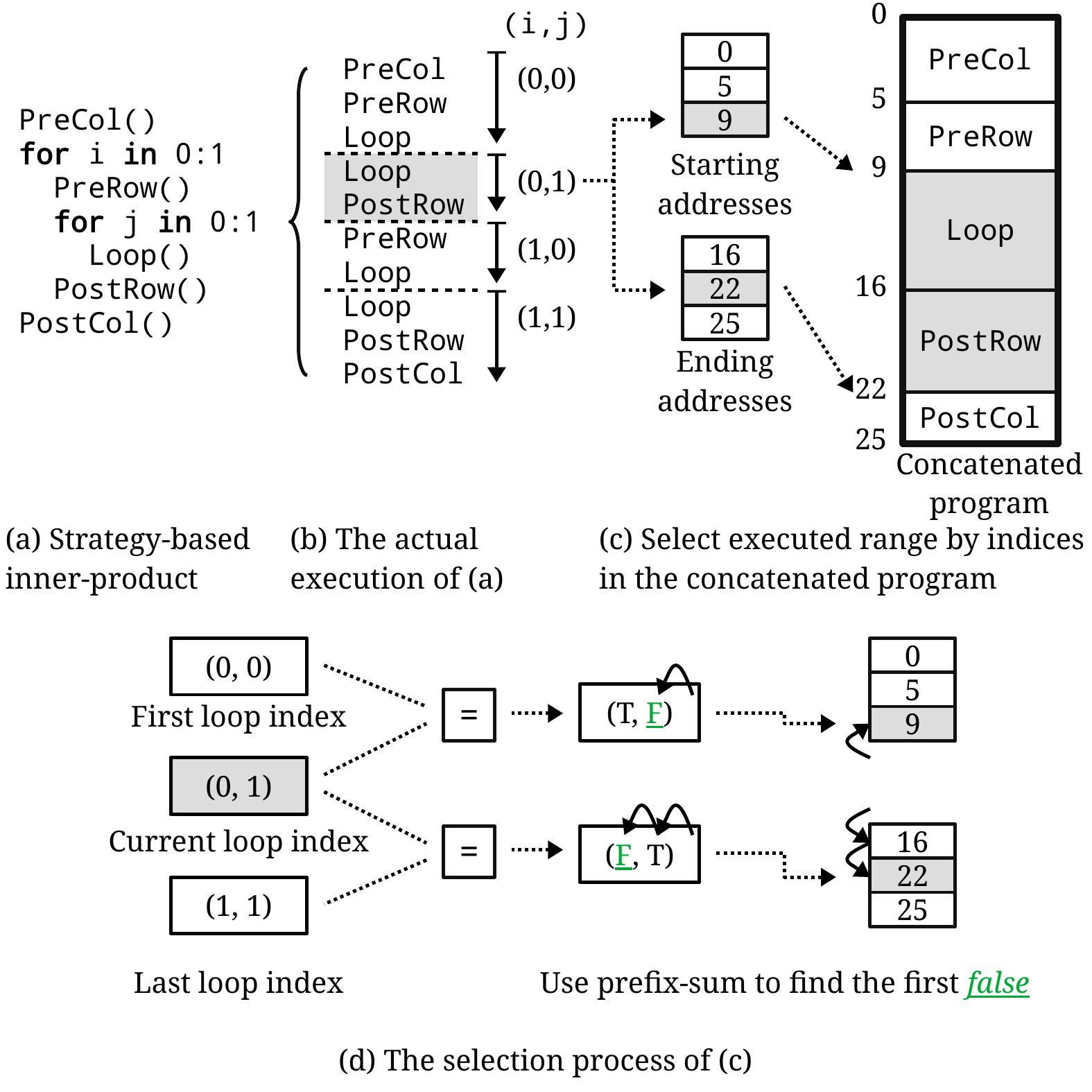}
\caption{\textbf{Ranged Inner-Product.} Execute a 2D tensor product by flattening ranges of a strategy class. It extends the strategy class of \lstref{inner_st} to the 2D case, so there are five strategy functions in (a) instead of three.}\label{fig:MERIT_inst}
\end{figure}

Extending the idea of strategy beyond vectors, programmers can use a more complex \textit{class} and nested \textit{for-loops} to implement a 2D Ranged Inner-Product, which is a $2\times 2$ for-loop.
As shown in \figref{MERIT_inst}a,
For 3D or higher dimensional tensors, however, this would not be suitable because of the resulting deeply nested \textit{for-loops}, and therefore we propose to linearize the nested loops.
\figref{MERIT_inst}b shows the function sequences being executed for the 2D case,
where each index corresponds to a certain execution order of the functions.
\figref{MERIT_inst}c and \ref{fig:MERIT_inst}d show the efficient Ranged Inner-Product implementation.
First, the strategy functions are concatenated to form a single program,
and their starting and ending addresses are stored as tables.
Given the loop indices (\ie \verb+i+, \verb+j+),
the starting and ending addresses are selected from the table,
which decides the function \textit{range} that should be executed for those indices.
To support higher-dimension execution,
we can simply enlarge the address tables and implement the selection using parallel prefix-sum.

\subsection{Putting It Together}\label{sec:pro:tensor}

In summary, the MERIT transform alters an input tensor into another tensor,
for the purpose of converting parallel algorithms into simpler parallel tensor reductions, or the Ranged Inner-Products.
This means that in the previous examples (\figref{merit_unrm} and \figref{merit_unrc}),
the matrices $M(\bA)$ and $M(\bB)$ are in fact tensors flattened into 2D matrices.
This flattening process is based on splitting the tensor indices into two parts.
The row indices $\bp$ reflect the algorithm \textbf{p}arallelism,
and column indices $\ba$ reflect the number of \textbf{a}ccumulated elements in a tensor product.
In these figures, the rows and columns of $M(\bA)$ and $M(\bB)$ are assigned with grid
indices like $(0,0)$, $(0,1)$, $(0,2)$, $(1,0)\cdots$.
We then use the notation $M(\bA)_{\bp,\ba}$ to refer to one element in a transformed tensor,
and for convenience, we denote the concatenated indices as $\mathbf{k} = (\bp,\ba)$.
Changing $k_j$, the $j$-th component of $\mathbf{k}$, which belongs to either $\bp$ or $\ba$,
is equivalent to moving along a particular axis $d_j$ of $\bA$
with a certain stride $s_j$ and offset $o_j$,

\begin{figure}[t!]
\centering
\subfloat[Tiled convolution.]{%
\includegraphics[width=0.48\textwidth,page=4]{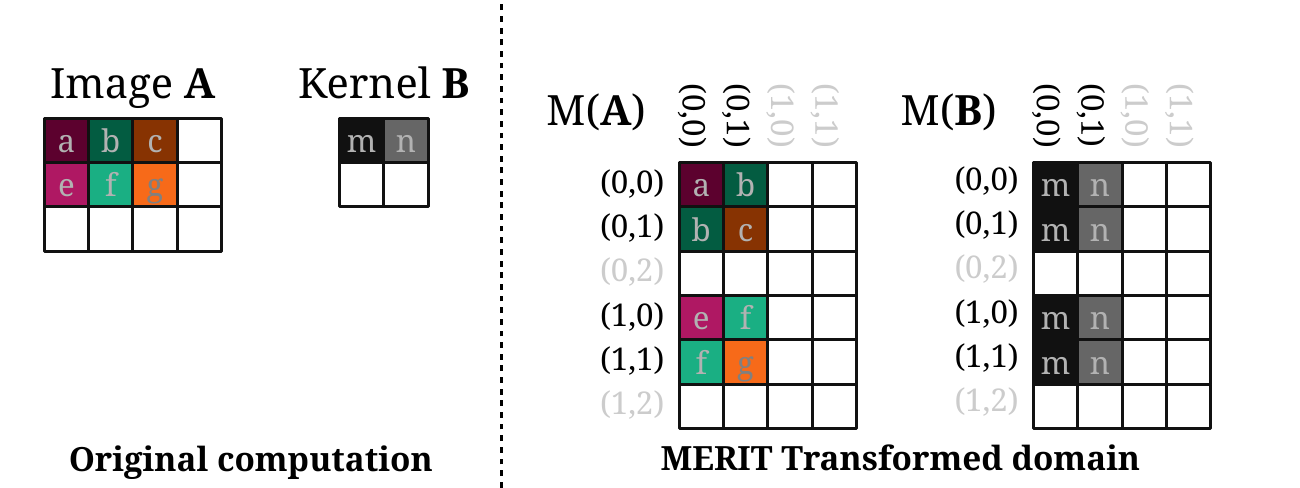}\label{fig:merit_unr_shmc}}\\
\subfloat[Tiled matrix multiplication.]{%
\includegraphics[width=0.48\textwidth,page=5]{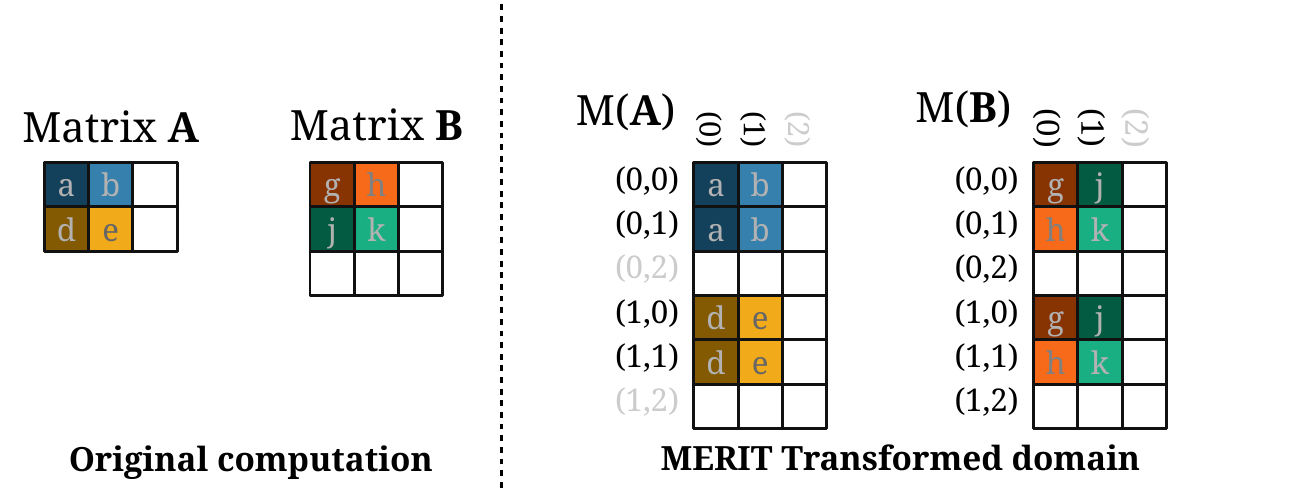}\label{fig:merit_unr_shmg}}
\caption{\textbf{MERIT transform with tiling.} Practical limits of computational resources dictate processing of subsets of the workload, which is equivalent to dividing the MERIT tensor into sub-tensors.}\label{fig:merit_unr_shm}
\end{figure}

\begin{equation}\label{equ:umi_um}
	M(\bA)_{\bp,\ba} = \bA_\bx \text{, where } x_i = \sum_j \delta_{i, d_j} (k_js_j+o_j),
\end{equation}
where $\delta$ is the Kronecker Delta.
The equation above first calculates the indices $\bx$ through $(\bp,\ba)$ and parameters $d_j, s_j, o_j$,
then $\bx$ is used to locate a specific element $\bA_\bx$ in $\bA$.

Based on the discussions above, a MERIT transformed tensor should reflect the total complexity and available parallelism of a computation.
For example, an $(m, k)$-by-$(k, n)$ GEMM problem has a total complexity of $\Theta(mnk)$ with $\Theta(mn)$ available parallelism, and the sizes of the transformed tensors are $((m,n),(k))$.
Similarly, a $(k\times k)$-convolution problem on an $(h\times w)$ image has a total complexity of $\Theta(hwk^2)$ with $\Theta(hw)$ available parallelism, and the sizes of the transformed tensors are $((h, w), (k, k))$.

We show an example for the MERIT transform of AlexNet~CONV1.
This layer adopts a stride size $4$ and a $11\times 11$ kernel size, so it can be written as
\begin{equation}\label{equ:alex_conv1}\begin{aligned}
&(p_1,p_2,p_3,a_1,a_2,a_3)\in\text{NDRange}(48,55,55,3,11,11)\\
&\left\lbrace\begin{aligned}
    M(\bI)_{p_1,p_2,p_3,a_1,a_2,a_3} &= \bI_{a_1,4p_2+a_2-5,4p_3+a_3-5}\\
    M(\bk)_{p_1,p_2,p_3,a_1,a_2,a_3} &= \bk_{p_1,a_1,a_2,a_3}
\end{aligned}\right.\,,
\end{aligned}\end{equation}
where $\bI$ and $\bk$ is the input feature map and convolution kernel.
Note that here we implicitly define the $x_i$ of \equref{umi_um} in the subscripts of right hand side expression.

Apart from CNN with stride, we can express several popular CNN variants with MERIT transform definition.
For example, the dilated CNN~\cite{dilconv} which can effectively increase the receptive field is defined by
\begin{equation}M(\bI)_{p_1,p_2,p_3,a_1,a_2,a_3} = \bI_{a_1,p_2+2a_2,p_3+2a_3}\,.\end{equation}
The correlation layer~\cite{flownet} is a CNN layer for optical flow.
It simulates the traditional motion estimation computation and can be expressed with
\begin{equation}\label{equ:corr}
\left\lbrace\begin{aligned}
    M(\mathbf{I1})_{p_1,p_2,p_3,p_4,a_1} &= \mathbf{I1}_{a_1,p_1,p_2}\\
    M(\mathbf{I2})_{p_1,p_2,p_3,p_4,a_1} &= \mathbf{I2}_{a_1,p_1+p_3,p_2+p_4}
\end{aligned}\right.\,.
\end{equation}

\section{Efficient MERIT Transform on GPUs}\label{sec:imp}
\begin{figure*}[ht]
\centering
\includegraphics[width=0.99\textwidth]{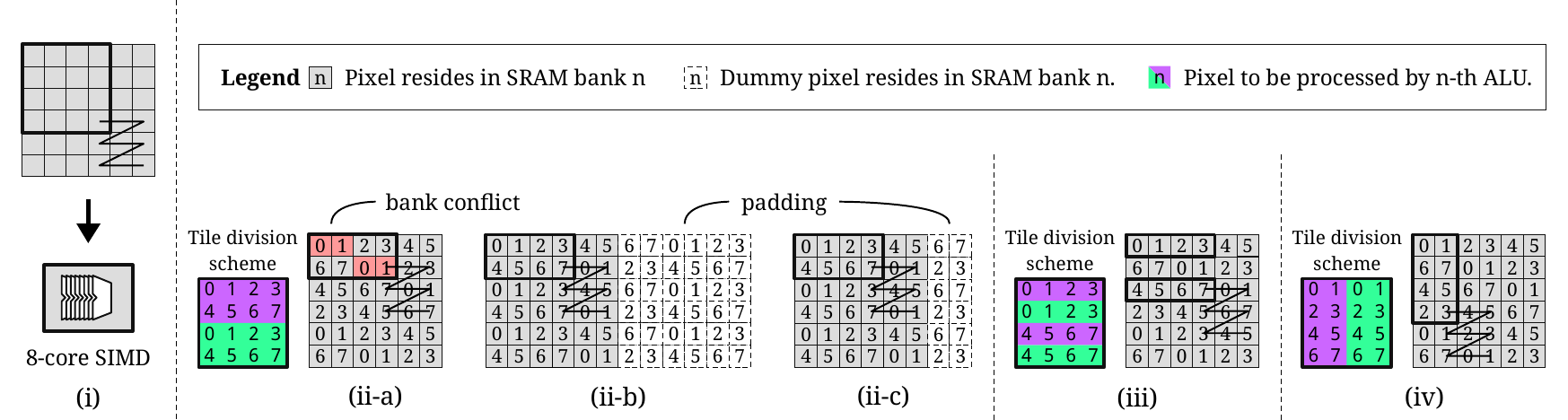}
\caption{\textbf{Bank conflict avoidance with MERIT.} (i) A $3\times 3$ convolution uses a block of $4\times 4$ threads with 8 cores, generating a $6\times 6$ patch in a 8-bank shared memory. (ii-a) A naive data layout causes conflict. (ii-b,c) Conflict-free data layouts. (iii,iv) Conflict-free thread grouping with re-tiling technique.}\label{fig:mex}
\end{figure*}

Modern GPUs have SIMD-like designs that are suitable for workloads targeted by the MERIT transform.
However, because of the intricate design of GPU memory hierarchy, care must be taken while fetching data, as the latency introduced by cache misses tends to be fairly high.
Traditionally, in order to maximize data locality, programmers are responsible for dividing workloads into independent sub-tasks and plan the data fetches accordingly.
These very tedious \textit{tiling} or \textit{blocking} optimization processes can be replaced by factorized MERIT transforms, as shall be explained in the remainder of the section.

\subsection{Tiled Parallel Execution}
With the MERIT transform, supporting \textbf{t}iled execution can be simply achieved by dividing $M(\bA)$ and $M(\bB)$ into sub-tensors of sizes $(\bt_p, \bt_a)$.
\figref{merit_unr_shm} highlights some sub-tensors from \figref{merit_unrc} and \figref{merit_unrm}
with $(\bt_p, \bt_a) = ((2,2),(1,2))$ and $((2,2),(2))$, respectively.
It can be observed that for an arbitrary tensor $\bA$ and its transformed tensor $M(\bA)$,
each sub-tensor of $M(\bA)$ can be fully contained by a minimal sub-tensor of $\bA$, whose size at the $i$-th dimension is calculated by this equation
\begin{equation}\label{equ:footprint}
	1+\sum_j (t_j-1)s_j\delta_{d_j,i},
\end{equation}
where $d_j$ and $s_j$ represent the same axes and strides as in \equref{umi_um},
and $t_j$ denotes the size at the $j$-th dimension of the sub-tensor in the $M(\bA)$ domain.
This allows us to compute the memory footprint of the sub-tensor.
For example, using a $5$-by-$5$ kernel in convolution in order to obtain a $16$-by-$8$ pixel output,
we assign $(\bt_p, \bt_a)=((16, 8), (5,5))$ in \equref{footprint},
which gives the memory size requirement of $(1+1(16-1)+1(5-1), 1+1(8-1)+1(5-1)) = (20,12)$.
Instead of loading all elements of $M(\bA)$ directly from $\bA$, we first load its sub-tensor into user-addressable shared memory, and programmatically transform the tensor to the $M(\bA)$ domain on-the-fly, thereby eliminating all memory bandwidth overhead caused by duplication, since all data are load from the shared memory.

\subsection{Bank Conflict Avoidance}\label{sec:proc:bank_conflict}

In GPUs, the shared memory may consist of many SRAM banks,
and the number of banks is usually designed to match the number of processor cores.
When the SIMD requests elements from the same SRAM at the same time,
the system performance drops drastically, causing the bank-conflict problem.
To avoid this problem, note that the vertical direction of the MERIT transformed matrices reflects the parallelism.
This property reduces bank-conflict problem into the analysis of column vector addresses within the sub-tensors of $M(\bA)$.

\figref{mex} illustrates examples for resolving bank-conflict
by padding~\cite[chap.~39]{gg3}, XOR-hash~\cite{xor1,xor2}, and our proposed \textit{re-tiling} technique.
The example shows a vector processor with 8 ALUs computing a $3\times 3$ convolution for a $4\times 4$ output block, requiring a $6\times 6$ block in total (\figref{mex}(i)).
The right parts of the figure illustrate some arrangements for distributing the $6\times 6$ pixel block into 8 SRAM banks physically.
In (ii-a), conflicts are highlighted in red,
whereas bank conflicts in (ii-b) are resolved by padding $6$ elements after every $6$ elements.
To reduce the padding waste, (ii-c) uses only $2$ padding with XOR-hash,
achieved by swapping the first and last four elements in every other row.
(iii) and (iv) show two different re-tiling that can create conflict-free scenarios without any hashing or padding requirement.
Later in \secref{proc}, we shall provide more details on finding suitable configurations and for storing and transferring of SRAM data.
We will also discuss our open-source implementation of the MERIT transform on GPUs in more details in \secref{eva}.

\section{Hardware Design for MERIT}\label{sec:proc}
\subsection{The MERIT-z Vector Processor}
\begin{figure}[t!]
\centering
\includegraphics[width=0.48\textwidth]{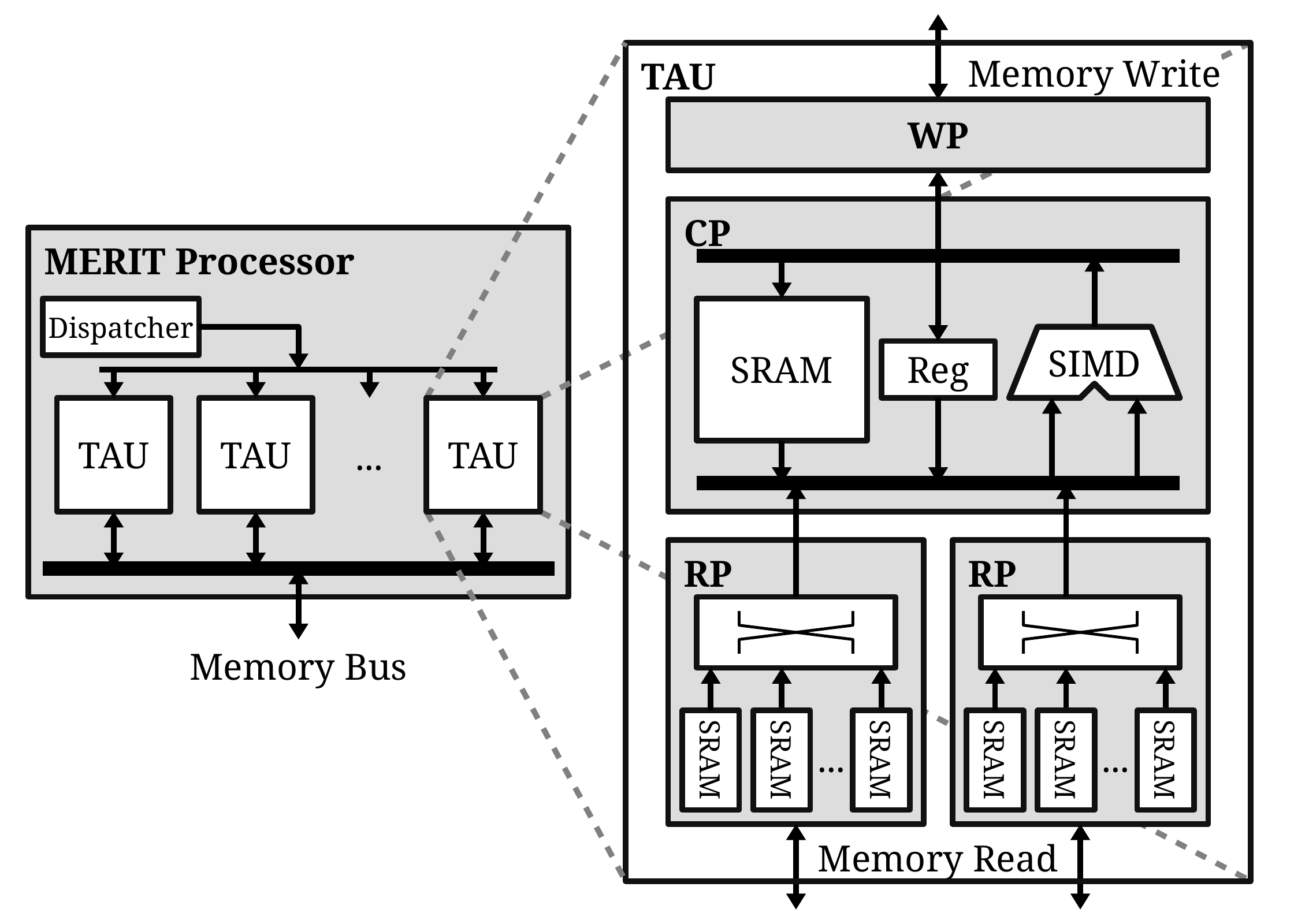}%
\caption{The MERIT-z processor architecture.}\label{fig:MERIT_proc}
\end{figure}

\begin{figure}[t!]
\centering
\includegraphics[width=0.49\textwidth]{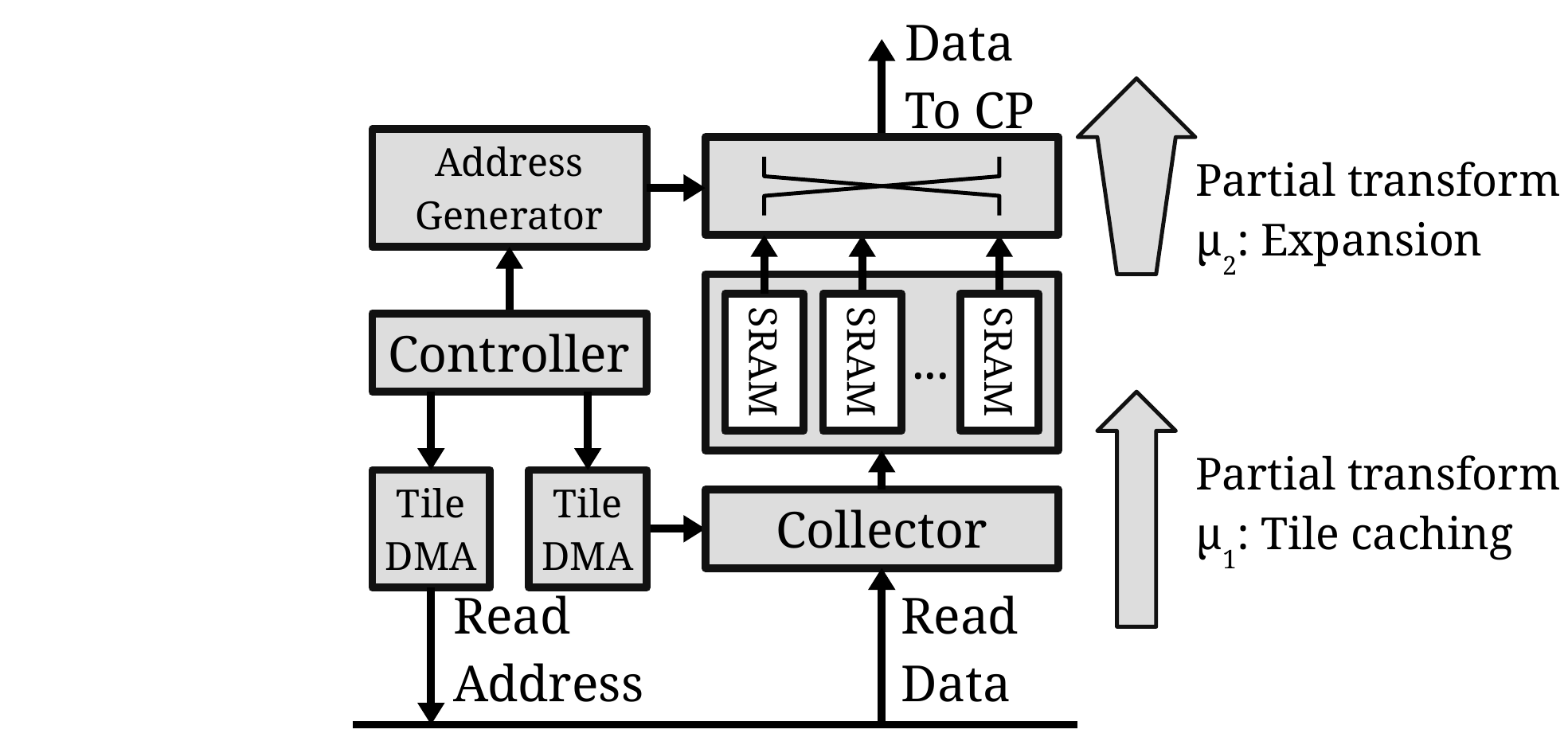}%
\caption{\textbf{The Read Pipeline (RP).} It executes sub-steps of a factorized MERIT transform, including two transforms representing data movement from DRAM through SRAM to SIMD. The tensors are expanded as late as possible to minimize memory requirement.}\label{fig:MERIT_rp}
\end{figure}

\begin{figure}[t!]
\centering
\includegraphics[width=0.49\textwidth]{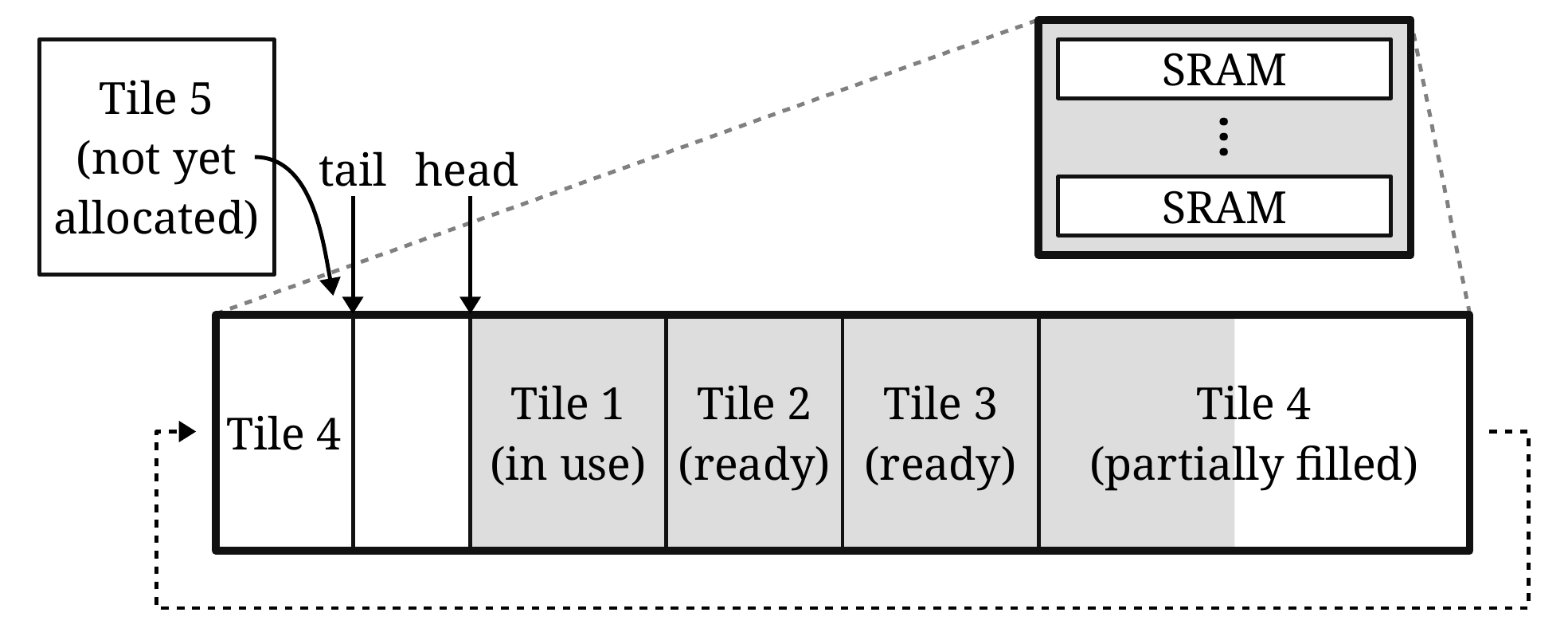}%
\caption{\textbf{Buffered tiles on SRAM.} The SRAM banks are used as one circular FIFO, and each tile is allocated, committed, and freed as an atomic unit.}\label{fig:circ}
\end{figure}

As shown above, GPUs combined with the MERIT transform framework can greatly simplify the optimization process for parallel algorithms.
However, GPUs also come with units that introduce unnecessary complexities in the memory hierarchy, leading to unpredictable performances.
In this section, we use the insights gained from the MERIT transform and create a clean-sheet ASIC design aimed to remove these complexities.
When done properly, it can outperform GPUs for memory-intensive vision processing applications while maintaining the same programming interfaces as the GPUs.

\figref{MERIT_proc} shows the proposed MERIT-z processor architecture.
It includes a MERIT Memory Management Unit (MMU) using classic circuit blocks such as the butterfly network to map common MERIT sub-steps onto the processor efficiently.
It also adopts a quite standard vector processor design,
which comprises of several \textit{Tile Accumulation Units} (TAUs) and one \textit{Dispatcher}.
A TAU is analogous to one Streaming Multi-Processors on GPUs, and is an array of $N=32$ ALUs cores in our implementation.
Our processor can be configured with variable numbers of TAUs,
such that it can be scaled for both low-power or high-end computing devices.
A TAU consists of three coarse-grained pipelines, namely the \textit{Read Pipeline} (RP), \textit{Compute Pipeline} (CP), and \textit{Write Pipeline} (WP).
The RPs cache tiles from the DRAM into the SRAMs, transform and feed them into the ALUs; the WP handles data that need to be written to the DRAM.
Together, RPs and WP form the MERIT MMU. 

\figref{MERIT_rp} shows the RP which consists of multiple SRAM banks and is designed to perform sub-steps of the MERIT transform.
The RP caches individual tensors tiles, according to \figref{merit_unr_shm} from the DRAM,
and performs the sub-step transform on-the-fly when the CP is comsuning the tiles.
This involves reading out the data from $N$ SRAM banks, shuffling, and feeding the data into $N$ cores.
The SRAM acts as a circular FIFO as shown in \figref{circ}.
2-port (TP) SRAMs should be adequate for this purpose,
or single-port SRAMs (SP) can be used in exchange for performance drop.
When the processor requests a tile, we allocate its corresponding space on the SRAM and then mark it as ready after receiving the entire tile from the DRAM.
Afterward, the CP can be launched after all dependent tiles are ready.
Meanwhile, tiles that are not needed anymore are released due to the nature of circular FIFOs.
Because partial tiles are not read out in this design,
the circuits handling the read-after-write hazards can be greatly simplified.
The total SRAM sizes in the two RPs of a TAU are $16$ and $8$~KB,
and they can provide enough bandwidth for many computation tasks.
For example, the RPs may hold the weights and feature maps in a CNN, two input matrices in GEMM,
or the reference and current frames in motion estimation.

\begin{table}[t]
\centering
\caption{The SIMD ISA of MERIT-z processor.}\label{tab:isa}
\begin{tabular}{llp{0.13\textwidth}}
\toprule
Operation type & Functionality & Possible Usage(s)\\
\midrule
Addition & \verb|a+((b+c)>>s)| & Common operation.\\\hline
Subtraction & \verb|a+((b-c)>>s)| & Common operation.\\\hline
1-norm & \verb|a+(abs(b-c)>>s)| & Clustering, motion estimation.\\\hline
MAC & \verb|a+((b*c)>>s)| & FC, CONV.\\\hline
Logical & \verb|max/min(a,b)| & ReLU, pooling.\\
 & \verb|a ? b : c| & \\
 & Binary bitwise ops & \\\hline
Indexing & Load a tensor index & \verb|meshgrid|, bilateral filter.\\\hline
Lookup & Lookup with interpolation  & Non-linear response in DNN, bilateral filter, or division.\\
\bottomrule
\end{tabular}
\end{table}

\begin{table}[t]
\centering
\caption{A brief summary of DNN hardware configurations.}\label{tab:alu_sram}
\begin{tabular}{lrrr}
\toprule
Hardware & SRAM/Word Size (KB/B) & \#ALUs & (\#ALUs/kword)\\
\midrule
NVIDIA 1080Ti          &  3584 / 4 &  3584 &  4.00\\
Eyeriss~\cite{eyeriss} &   192 / 2 &   168 &  1.76\\
TPU~\cite{tpu}         & 28000 / 1 & 65536 &  2.34\\
DianNao~\cite{diannao} &    48 / 2 &   256 & 10.66\\
MEARI~\cite{maeri}     &   164 / 2 &   168 &  2.05\\
Ours (Per TAU)         &    29 / 2 &    32 &  2.20\\
\bottomrule
\end{tabular}
\end{table}

The CP accepts data from the RPs, executes the Ranged Inner-Products, and yields data to the WP.
A CP contains a SIMD array and a 5~KB SP~SRAM for storing partial results.
The SIMD ALUs run a 32b-ISA comes with only seven kinds of instructions (\tabref{isa}) for 16b fixed-point arithmetic operations, but can support a wide range of applications.
The simplicity of this ISA is enabled by offloading the complexity of data movement and address calculation to the MERIT MMU.

The WP does not store any data.
Instead, it shuffles and collects data from CP by generating addresses on-the-fly assembling the output lines to aligns with the DRAM or cache lines.
Since writing data does not cause any execution dependency, a TAU can never stall at WP unless the DRAM write queue is full.

We select the total sizes of SP SRAM in the two RPs and the CP according to several empirical rules.
First, \tabref{alu_sram} shows the total SRAM sizes and ALU number of several popular DNN architectures.
While these architectures come with such variant computation ability and datapath,
their SRAM and ALU ratio are close, and we also select a ratio in line with them.
Second, the two buffers in RPs are not sharable in our architecture,
and we find that the buffered tensor sizes in two RPs are mostly unequal.
For example, in motion estimation or \cite{flownet}, the buffered tensors of the searched frame are always larger than the those of the current frame.
We thus believe that using asymmetric buffer sizes provides better utilization.
Third, as we discuss in \figref{MERIT_rp}, the partial sum is stationery in the buffer while the buffers in RPs are also used for prefetching.
This suggest that the buffer in RPs should be a few times larger than that in the CP.
Based on these observations, we come up with the $16$, $8$~KB, and $5$~KB configuration.

Overall, a MERIT-z processor has two external interfaces, namely, 
a data interface which uses the valid/ready channel similar to the standard buses like AXI~\cite{AXI} for both address and data, and 
a control interface for transmitting the transform parameters $d,s,o$ in \equref{umi_um} and the program shown in \figref{MERIT_inst}.
Unlike many accelerators that tend to consume tens of KBs of instruction memory to work,
our processor only uses less than 0.5~KB of registers to define a computation kernel.

\begin{figure}[t]
\centering
\includegraphics[scale=0.4]{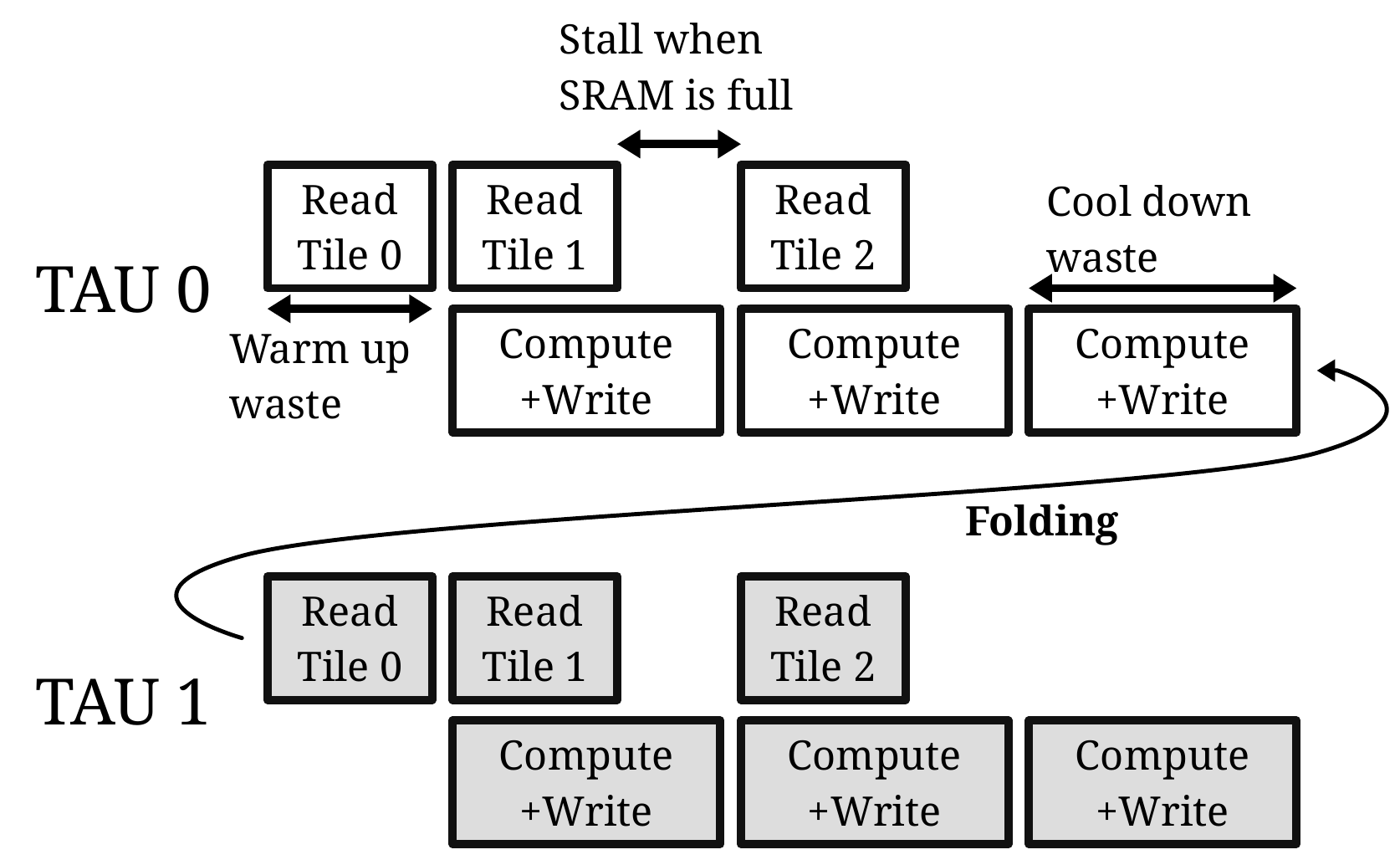}
\caption{\textbf{Overlapping computation and prefetching.} The MERIT-z processor supports overlapped data movement and computation intrinsically.  We can also adopt \textit{folding} to eliminate wastes.}\label{fig:prefetch}
\end{figure}

While most of the memory related optimization is encoded in the MERIT transform,
further optimization on MERIT-z pipeline can be done by overlapping computation and memory fetching as shown in \figref{prefetch}, since these two operations use different hardware units in a TAU.
To further eliminate bubbles during warm-up and cool-down stages of the pipeline, a common technique is to reduces the parallelism by forcing independent works to run on the same TAU, which is also known as \textit{folding}.
Instead of creating dedicated hardware units, these pipelining techniques can be implemented entirely in software when algorithms are expressed in the MERIT transform by merely doubling the width and halve the height of $(M(\bA), M(\bB))$, and add an extra \textit{for-loop} level in the Ranged Inner-Products.

\subsection{Architectural Comparison with DNN Accelerators}\label{sec:proc:archcomp}

\begin{figure*}
\centering
\subfloat[Systolic, $8\stimes16$ ALUs, no input local buffer (ILB)]{\frame{\includegraphics[scale=0.36,trim={0 0 4.25cm 0},page=1]{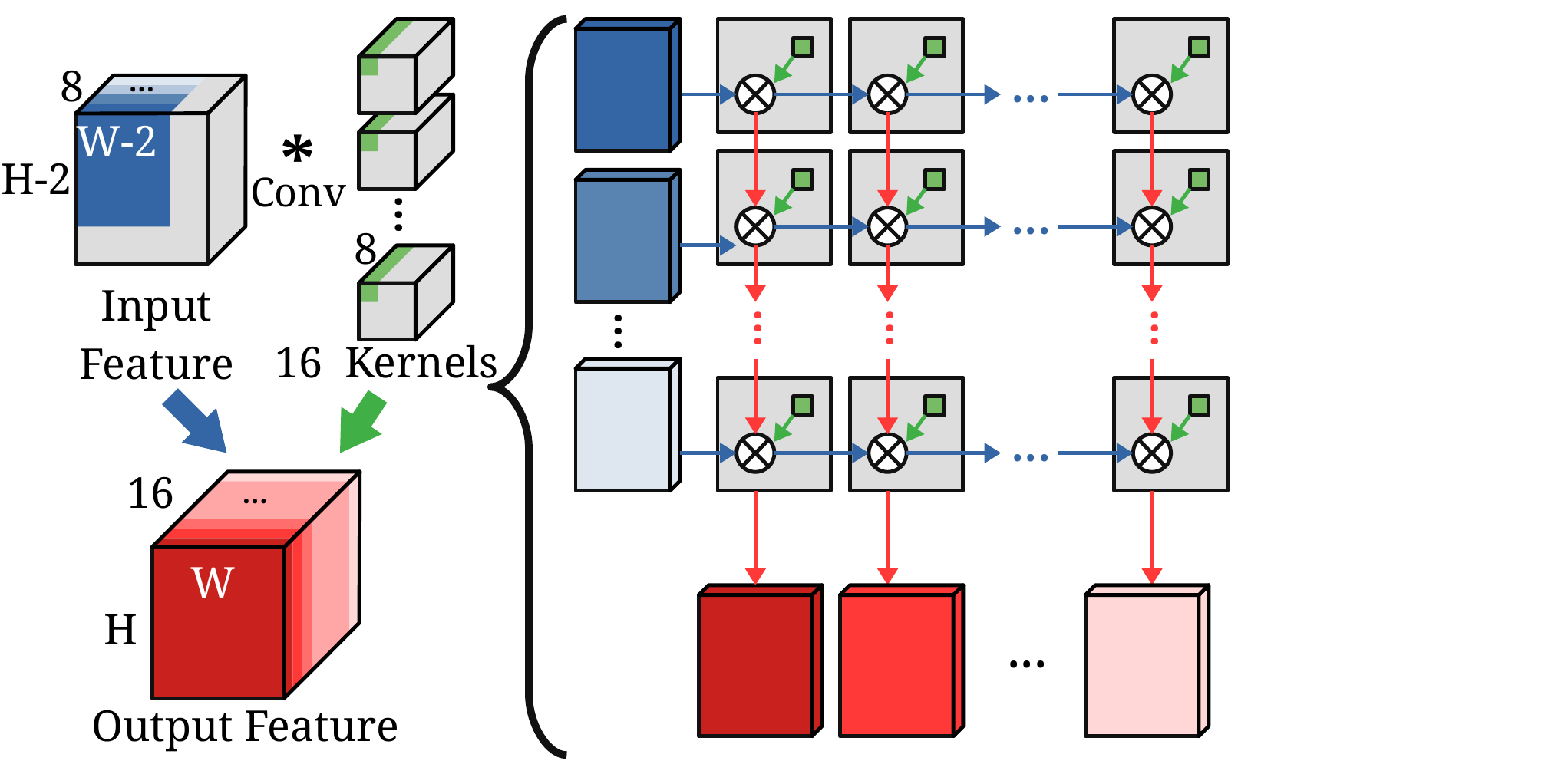}\label{fig:taxonomy:systolic}}}\hspace*{1em}%
\subfloat[Eyeriss, $3\stimes13$ ALUs, $19.5$~KB ILB]{\frame{\includegraphics[scale=0.36,page=2]{taxonomy_v2.pdf}\label{fig:taxonomy:eyeriss}}}\hspace*{1em}%
\subfloat[MERIT-z, $32$ ALUs, $24$~KB ILB]{\frame{\includegraphics[scale=0.36,trim={0 0 8.5cm 0},page=3]{taxonomy_v2.pdf}\label{fig:taxonomy:merit}}}%
\caption{\textbf{Comparison of data reuse in DNN accelerators.} Compared with the systolic array which is for GEMM, accelerators like Eyeriss duplicate data locally to improve data reuse. MERIT-z processor distributes a subtensor across local SRAMs and duplicates data through the butterfly network to ensure high data reuse.
The validness of the butterfly network is mathematically guaranteed in the form of a MERIT transform tensor in \secref{propose:butterfly}.}\label{fig:taxnomy}
\end{figure*}

\begin{figure}[t]
\centering
\includegraphics[scale=0.34]{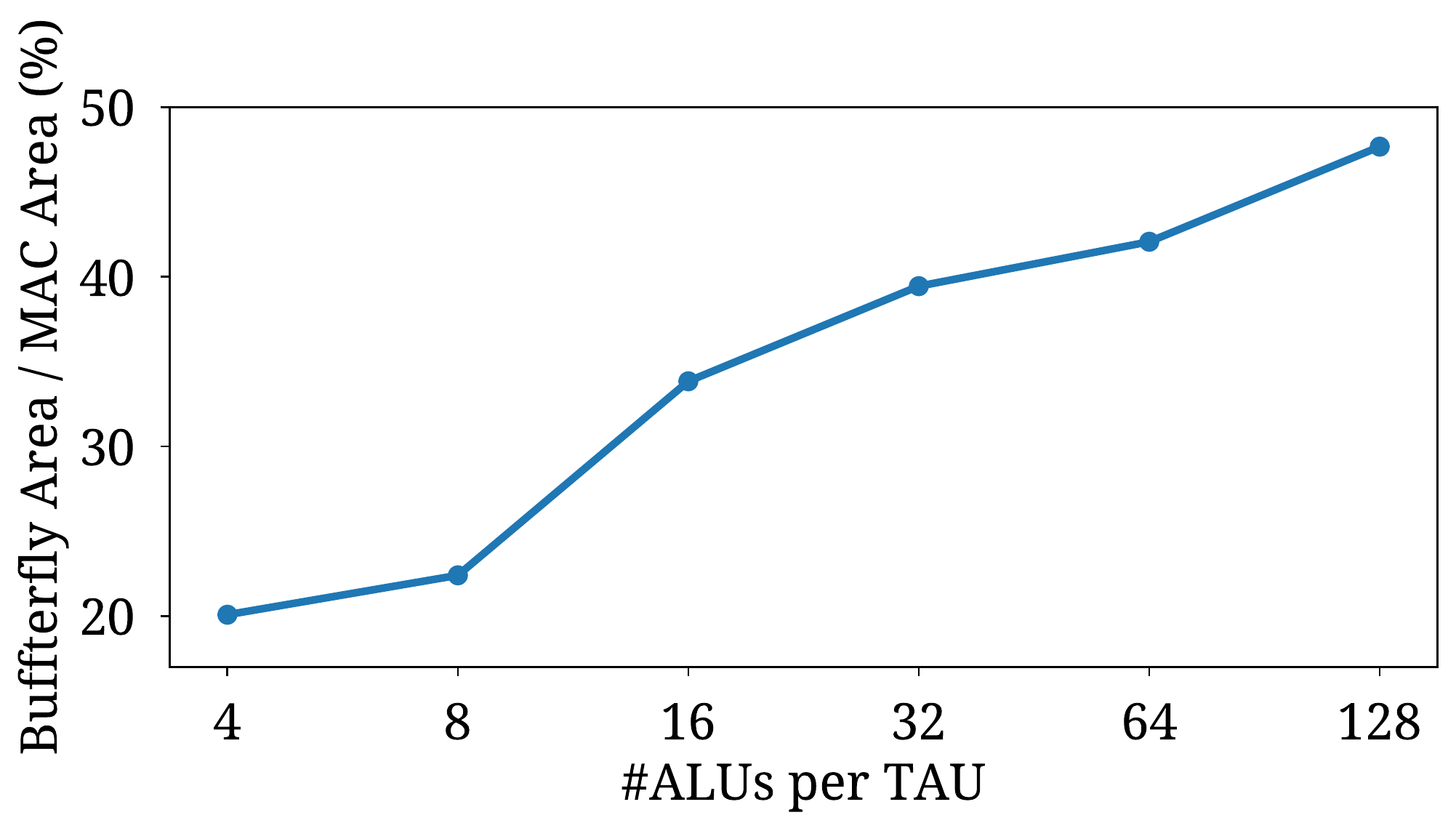}
\caption{\textbf{Area overhead of the butterfly networks.} According to the results, there are $32$ ALUs in a TAU, and the two $32$-to-$32$ butterfly networks cost $40$\% of the $32$ MAC units area.}\label{fig:analyze_bf}
\end{figure}

\begin{figure}[t]
\centering
\includegraphics[scale=0.34]{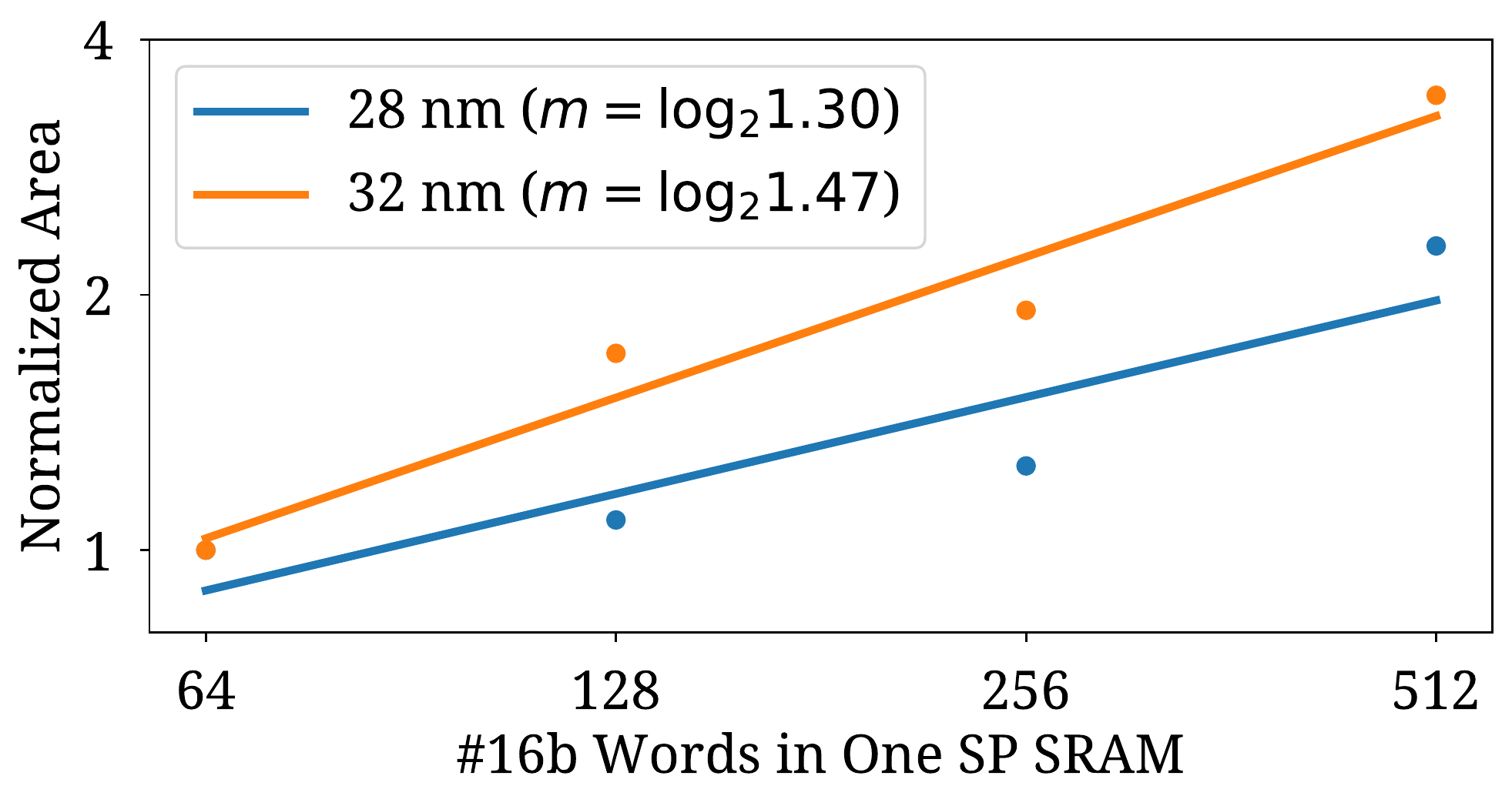}
\caption{\textbf{Normalized SRAM area for local buffer.} For the local buffers in most DNN accelerators, a doubled storage capacity only leads to a $1.30$-$1.47$x area.}\label{fig:analyze_sram}
\end{figure}

\begin{figure}[t]
\centering
\includegraphics[scale=0.43]{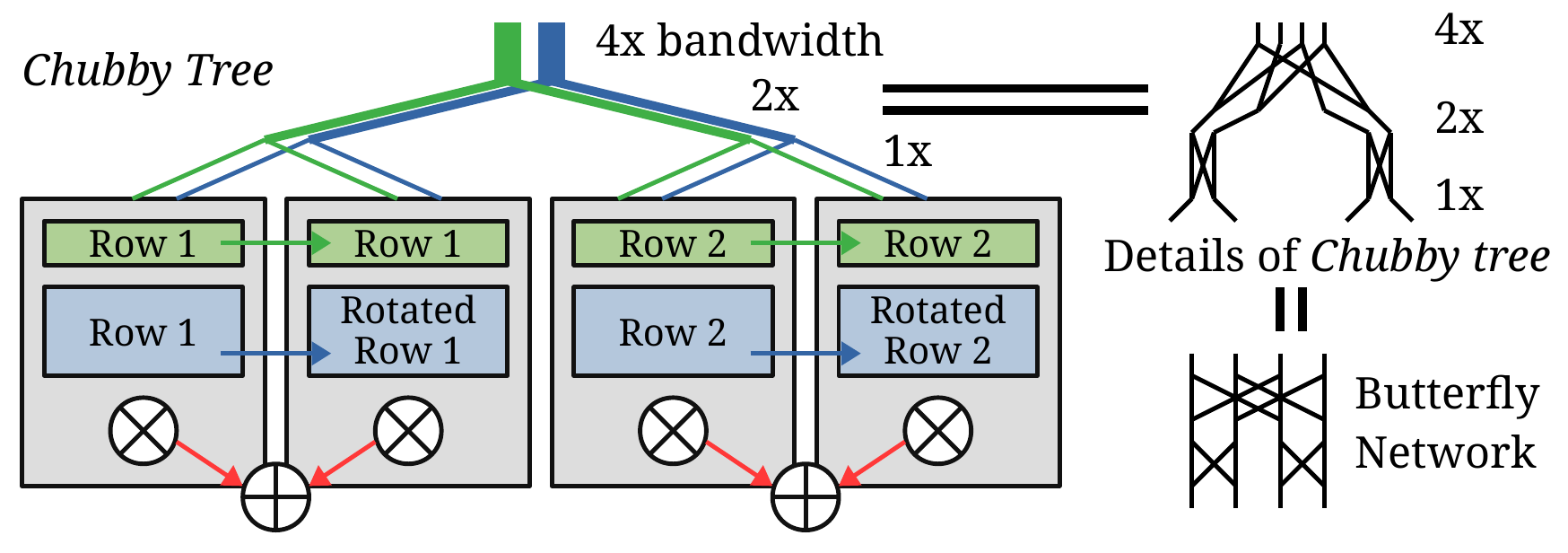}
\caption{\textbf{Butterfly network used by MAERI.} MAERI connect all its ALUs with the tree-like connection. It distributes data to every local buffer through Chubby Tree, which is essentially equivalent to the traditional buffer network used by MERIT-z.}\label{fig:maeri_bf}
\end{figure}

\begin{table}[t]
\setlength\tabcolsep{2pt}
\centering
\caption{A comparison of ALU bandwidth in one processing pass of a $3\times3$ CNN workload.}\label{tab:arch_reuse}
\begin{tabular}{@{}lrrrrr@{}}
\toprule
\footnotesize
Architecture & Input   & Kernel & Output  & MAC & Reuse\\
(shape)      & feature &        & feature &     & rate\\
\midrule
Systolic $(8\stimes16)$\\
\hspace*{1em}1~ALU & $1$ & $1$ & $1$ & $1$ & $0.33$\\
\hspace*{1em}Overall & $8n$ & $128$ & $16n$ & $128n$ & $5.33$\\
\hline
Eyeriss $(3\stimes14)$\\
\hspace*{1em}1~ALU & $3\stimes4$ & $16\stimes3\stimes4$ & $16$ & $16\stimes3\stimes4$ & $0.87$\\
\hspace*{1em}1~Pass & 1~ALU$\stimes16$ & 1~ALU$\stimes3$ & 1~ALU$\stimes14$ & 1~ALU$\stimes3\stimes14$ & $8.12$\\
\hspace*{1em}Overall & 1~Pass$\stimes n$ & 1~Pass & 1~Pass$\stimes n$ & 1~Pass$\stimes n$ & $19.38$\\
\hline
MERIT-z $(32)$ & $18\stimes10\stimes8$ & $3\stimes3\stimes8\stimes16$ & $0$ & $3\stimes3\stimes8\stimes16\stimes8\stimes16$ & \textbf{78.77}\\
\bottomrule
\end{tabular}
\end{table}

The MERIT-z processor is more general than dedicated DNN processors, but we can achieve comparable or better performances compared to these processors like the systolic array~(TPU) or Eyeriss.
We attempt to provide some insights by comparing the dataflow on an architectural level.
As shown in \figref{taxnomy}, we illustrate a $3\times3$ CNN workload example,
assuming the following architectures: (1) a $16\times 8$ systolic array, (2) a $12\times 14$ Eyeriss divided into four $3\times 14$ groups (as described in their paper), and (3) MERIT-z processor with four $32$-ALU TAUs.
In \tabref{arch_reuse}, we analyze the data-reuse rate of these architectures,
which is defined by the MAC count divided by the input and output word count.
Among all architectures, MERIT-z has a significantly higher reuse rate than the others.
This is why the current MERIT-z implementation does not include an extra memory hierarchy~(\ie global buffer) between the TAUs and DRAM.

Now we take a closer view of the dataflow of these architectures.
The systolic array is an architecture for matrix multiplication, so we must use $9$ processing passes to compute the convolution~(\figref{taxonomy:systolic}).
Every cycle, every ALU loads two inputs, performs a MAC and delivers one output value to its neighbor.
Overall, each pass inputs an $8n$ matrix and outputs a $16n$ matrix, performing $128n$ MACs in total.
This gives the per-ALU reuse rate $1/(1+1+1)=0.33$ and overall reuse rate $128/(16+8)=5.33$.
Since systolic array has no local~(\ie per-ALU) storage, the local reuse rate is only $0.33$.

Eyeriss is a systolic array variant with extra local storage to improve the reuse rate.
Every Eyeriss ALU can hold up to $3\times4$ input feature maps, $3\times4\times16$ kernels, and $16$ partial sums, performing $3\times4\times16$ MACs~(\figref{taxonomy:eyeriss}).
This almost triple the per-ALU data-reuse.
Besides, due to their application of the multi-cast network,
Eyeriss improves the overall reuse rate by $8.12/5.33-1 = 52\%$ with only $42$~ALUs and extra $21$~KB local storage, compared with the $128$-ALU systolic array;
this value can be further boosted to $19.38$ by their row stationary dataflow.

An apparent problem of the Eyeriss dataflow is the data duplication in local storage.
While their multi-cast connection efficiently broadcasts the kernels to a row of ALUs,
it also reduces the equivalent local buffer size by an order.
On the other hand, while MERIT-z also uses similar input local buffer sizes~(less than $1$~KB each),
we aggregate them into a large one with the butterfly network.
We prove in \secref{propose:butterfly} that the butterfly network can provide full throughput for all supported workloads.
The aggregated buffer sizes reach $8$--$16$~KB, allowing MERIT-z to perform a smaller but complete CNN workload~(\figref{taxonomy:merit}),
which enables a partial CNN workload in one TAU with a reuse rate $78.77$ as calculated in \tabref{arch_reuse}.
This makes MERIT-z more like a no-local-reuse~(NLR) architecture in Eyeriss taxonomy.
NLR architectures read all data from a larger global buffer and thus achieve a higher ALUs density and data reuse rate.
However, their frequent access to the large global buffers implies high area and power overhead efficiency compared to spatial architectures.
Since MERIT-z uses the butterfly network as its datapath,
the overhead is rather low compared with classic NLR architectures.
\figref{analyze_bf} is our synthesizing results illustrating that in a TAU with $32$ ALUs,
the two butterfly networks cost only $40$\% of $32$ MAC area.
Therefore, we can simultaneously benefit from the high data reuse of NLR architecture as well as the power-efficiency of systolic-like architectures.
The efficiency of butterfly networks is also demonstrated in accelerators like MAERI~\cite{maeri} as shown in \figref{maeri_bf}.
It uses a \emph{Chubby Tree} to distribute data from the global buffer to local buffers and collects the partial sum through a configurable adder tree.
The Chubby Tree in MAERI provides higher flexibility, but it still has data duplication in its local buffers.
Besides, the chubby tree and the classic butterfly network are essentially topologically equivalent.

MERIT-z has several design advantages in its local storage over the other architectures,
and it also has advantages owing to its simplicity of memory hierarchy.
First, the partial sums are written to the DRAM only when the summation tasks are finished, which satisfies the output stationary data reuse, so the output bandwidth is zero in \tabref{arch_reuse}.
Second, since the ALUs are organized as a SIMD that executes synchronously,
the ALUs always issue vector read and write,
and we use a wide SRAM to reduce the amortized partial sum buffer area.
Last, since we can use the same input buffer for both prefetching data and supplying data for SIMD~(\figref{circ}),
we can remove the requirement for a dedicated FIFOs that costs $30$\% of the local buffer area in Eyeriss.
For $28$--$32$~\si{nm} SP SRAMs less than $1$~KB, a $1.30$--$1.47$x area means doubled storage capacity~(\figref{analyze_sram}),
and this explains why we can use a larger per-ALU \emph{input} local buffer size~($0.75$~KB) than Eyeriss~($0.50$~KB).

\subsection{Efficient Memory Distribution Circuit with MERIT}\label{sec:propose:butterfly}

In \ssecref{proc}{bank_conflict}, we mention that each ALU needs an SRAM bank in order to maximize processor efficiency and avoid bubbles.
A static assignment between ALUs and SRAMs can be too inflexible, but a crossbar with $\Theta (N^2)$ multiplexers would be too power-hungry and complex.
In our processor, we find that the classic butterfly network works well with MERIT transform while utilizing only $\Theta (N\lg N)$ multiplexers,
which represents only 30\% of the area compared to a full crossbar in our synthesized circuits.
In the remainder of the section, we show that, for computer vision algorithms,
how a butterfly network can correctly shuffle elements from SRAM banks to ALUs without stalls using the MERIT transform.

Given the access patterns shown earlier in \figref{mex},
we denote the SRAM address used by the $n$-th ALU as $A_n$,
and a data element lies in $A_n$ belongs to SRAM bank $(A_n \mod 8)$.
In the following discussion, we show the sufficient conditions for hardware correctness and efficiency of MERIT transform implementation.
Specifically, we observe that when mapping MERIT transform to SRAM and circuits, no bank conflict can occur, and a butterfly network can distribute the data to ALUs.
In \figref{mex}(ii-a), the addresses accessed by the first purple sub-tile are $A_{0:7} = (0,1,2,3,6,7,8,9)$.
Since the sub-tile size is the same as the ALU number,
the addresses can be formulized as:

\begin{equation}
A_n = A_0 + \sum_{i=0}^{2} c_i b_{n,i},
\end{equation}
where $b_{n,i}$ is the $i$-th bit ($b_0$ is LSB) of the binary representation of ALU index $n$.
For instance, for (ii-a), (ii-b), (iii), and (iv),
the $c_{0:2}$ are $(1,2,6)$, $(1,2,12)$, $(1,2,12)$, and $(1,6,12)$, respectively.
\equref{bin_rep} shows a binary representation of the addresses when $A_0 = 3$ and $c_{0:2} = (1,6,12)$, 
\begin{equation}\label{equ:bin_rep}
\begin{aligned}
A_{0:7} \mod 8
&=
\setlength\arraycolsep{2pt}
\begin{bmatrix}
3&4&1&2&7&8&5&6
\end{bmatrix}\\
&=
\setlength\arraycolsep{2pt}
\begin{bmatrix}
1&2&4
\end{bmatrix}
\begin{bmatrix}
1&0&1&0&1&0&1&0\\
1&0&0&1&1&0&0&1\\
0&1&0&0&1&0&1&1
\end{bmatrix}\\
&\equiv
\setlength\arraycolsep{2pt}
\begin{bmatrix}
1&2&4
\end{bmatrix} \bS
\end{aligned}.
\end{equation}
Upon careful inspection of this equation, we find that the matrix $\bS$ exhibits some special attributes that can be represented by a \textit{hash property matrix} $\bH$ whose elements $h_{i,j}$ can be computed using the matrix $\bS$, 
\begin{equation}
h_{i,j} =
\left\lbrace
\begin{aligned}
	0:\quad& s_{n,j} = s_{n\wedge 2^i,j} \quad \forall n\\
	1:\quad& s_{n,j} \neq s_{n\wedge 2^i,j} \quad \forall n\\
	x:\quad& \text{Otherwise}\\
	       & \ie (s_{n,j}, s_{n\wedge 2^i,j})\text{ are not constrainted}
\end{aligned}
\right. ,
\end{equation}
where $\wedge$ means the bitwise XOR function, and $n\wedge 2^i$ means flipping the $i$-th bit of n.
Readers can verify that we can obtain this \textit{property matrix} for \equref{bin_rep} regardless of the base addresses of $A_0$, 
\begin{equation}
\bH =
\setlength\arraycolsep{2pt}
\begin{bmatrix}
1&0&0\\
x&1&0\\
x&x&1
\end{bmatrix}.
\end{equation}
The positions of $1$s in $\bH$ are directly related to the power of $2$ of the prime factorization of $c_{0:2}$.
In the following discussions, we define the operators on symbols $(0,1,x)$ as the standard operators in \textit{ternary logic}.
Equipped with this hash property matrix, we assert that the sufficient condition for a MERIT transform to map to a butterfly network is whether the matrix $\bH$ can be reduced to the identity matrix $\bI$ as follows.
First, we select a row that contains only $0$s and $1$s.  Then, we compute the NOT version of this row and apply it to another row with the AND operation.
The process is repeated until it produces, or fails to produce, an $\bI$.
This reduction algorithm is similar to a standard Gaussian elimination process without row swapping.
For example,
\begin{equation}
\setlength\arraycolsep{2pt}
\bH_1 = 
\begin{bmatrix}
1&0&0\\
x&1&0\\
x&x&1
\end{bmatrix}
\end{equation}
represents compatible address mapping for the butterfly network, while
\begin{equation}
\setlength\arraycolsep{2pt}
\bH_2 = 
\begin{bmatrix}
1&0&x\\
x&1&0\\
0&x&1
\end{bmatrix}
\end{equation}
does not since we cannot find a row without $x$ to start.

In a more complex computation task, such as CNN with strides or dilated CNN, as shown in \figref{mex}(ii-c),
$\bH$ could become a nonsquare matrix.
When this happens, we apply the following procedure to map a nonsquare $\bH$ matrix into a square one $\bH'$, 

\begin{equation}\label{equ:bin_rep4}
\begin{aligned}
\bH' \equiv \bR\bX\bH &=
\setlength\arraycolsep{2pt}
\underset{\text{Bit rotate}}{\begin{bmatrix}
0&1&0\\
0&0&1\\
1&0&0
\end{bmatrix}}^1
\underset{\text{XOR hash}}{\begin{bmatrix}
1&0&0&0\\
0&1&1&0\\
0&0&1&1
\end{bmatrix}}
\begin{bmatrix}
0&0&1\\
0&0&x\\
1&0&x\\
x&1&x
\end{bmatrix}\\
&=
\setlength\arraycolsep{2pt}
\begin{bmatrix}
1&0&x\\
x&1&x\\
0&0&1
\end{bmatrix}
\end{aligned},
\end{equation}
where addition and multiplication are defined as XOR and AND operators in the ternary logic.
Equation \equref{bin_rep4} shows a $4\times 3$ property matrix $\bH$ created using $c_{0:2} = (4,8,3)$,
Two additional matrices, $\bR$ and $\bX$, are introduced to convert it into a squared matrix $\bH'$.
$\bX$ is an upper triangular matrix that trims $\bH$ into a square matrix,
and it has no more than one off-diagonal term per row.
Matrix $\bR$ is a row permutation matrix for swapping upper rows to the bottom, and it can be applied multiple times.
Again, readers can verify that the resulting matrix $\bH'$ fulfills the sufficient condition above.

Since matrices $\bS$, $\bR$, and $\bX$ are all binary matrices, we can map the math above into pure data shuffle circuits and place them in the Read Pipeline (\figref{MERIT_rp}).
For a $32$-core TAU, the data comes into the SRAM through the collector with a $\log_2 32 = 5$-stage butterfly network, followed by a $\log_2 5 = 3$-stage omega network that implements the matrices $(\bX, \bR)$ above.
The data move from the SRAM into the Compute Pipeline with another $\log_2 32 = 5$-stage butterfly network.
The discussions above imply that, we can use classic data distribution circuit blocks to effectively feed data into the processors to support a wide variety of workloads.

\section{Evaluation}\label{sec:eva}
\subsection{Code Size Reduction with MERIT}
In this experiment, we collect fast kernels from open-source projects such as Caffe, OpenCV, and Parboil~\cite{caffe,opencv,parboil}, and rewrite them in order to demonstrate the code size reduction effect with the MERIT transform.
Listing \ref{lst:bilat} and \ref{lst:bilat_call} show an example of the bilateral filter~\cite{bilateral} implemented as an inner-product strategy class,
where the Gaussian weights of spatial kernels are precomputed as lookup-tables.
In \verb+Loop()+, the neighbor pixels and two spatial kernels are packed in the array \verb+i+.
\lstref{bilat_call} specifies the input as an $h$-by-$w$ image divided into $16$-by-$16$ blocks, and each thread with a block performs a $k$-by-$k$ local window scan.
The middle lines represent the $M(\cdot)$s, and the range term $\sigma_r$ is passed to the kernel in the last line.

\begin{minipage}{\linewidth}
\begin{lstlisting}[language=c++, caption={A strategy class for bilateral filter.}, label=lst:bilat]
class BilateralStrategy {
   float wsum, wxsum, center;
   struct Constant {float norm;};
   // Note: Function signatures are macros.
   PreLoop(0,1) {wsum = wxsum = 0; center = i[0];}
   Loop(0,3) {
      float d = center-i[0];
      float w = expf(d*d*c.norm)*i[1]*i[2];
      wsum += w; wxsum += w*i[0];
   }
   PostLoop(1,0) {o[0] = wxsum/wsum;}
};
\end{lstlisting}
\end{minipage}

\begin{minipage}{\linewidth}
\begin{lstlisting}[language=c++, caption={Bilateral filter using MERIT transform.}, label=lst:bilat_call]
InnerProduct<BilateralStrategy>(
   // size and tile size
   {{h,16}, {w,16}}, {{k,k}, {k,k}},
   // d_i, o_i, s_i for input
   {{0,0,1}, {1,0,1}},
   {{0,-k/2,1}, {1,-k/2,1}},
   {h, w} // h-by-w image
   // d_i, o_i, s_i for spatial weight
   {/*Nothing*/},
   {{0,0,1}, {/*Nothing*/}},
   {k} // vector of length k
   // More configurations of M()s ...
   Bilateral::Constant{0.1}
);
\end{lstlisting}
\end{minipage}

\begin{table}[t]
\centering
\caption{Code token count comparison.}\label{tab:token}
\begin{tabular}{llrrr}
\toprule
    Kernel & Token Type & \multicolumn{1}{l}{MERIT} & \multicolumn{1}{l}{N\"aive} & \multicolumn{1}{l}{Open-Source} \\
                       && \multicolumn{1}{l}{on GPU} & \multicolumn{1}{l}{on CPU} & \multicolumn{1}{l}{on GPU} \\
\midrule
Motion     & Identifier & 49 & 80 & 194\\
Estimation & Operator   & 11 & 45 & 106\\
\hline
Bilateral        & Identifier & 69 & 87 & 122\\
Filter           & Operator   & 22 & 49 &  61\\
\hline
Forward     & Identifier & 53 & 110 & 204\\
Propagation & Operator   &  7 &  65 & 117\\
\hline
GEMM        & Identifier & 34 & 34 & 146\\
            & Operator   &  7 & 20 &  35\\
\hline
Integral & Identifier & 24 & 23 & 50\\
Image    & Operator   &  6 & 14 & 22\\
\hline
Separable & Identifier & 35 & 50 & 91\\
Filter    & Operator   &  5 & 29 & 47\\
\bottomrule
\end{tabular}
\end{table}

\tabref{token} lists the number of lexical tokens required to express different algorithms with MERIT, our n\"aive C++ implementation, and fast open-source kernels for the GPUs.
As shown in the table, because code regarding data movement is no longer needed in MERIT,
we can greatly reduce the number of both arithmetic operators and identifier tokens even compared to the na\"ive CPU implementations.

\subsection{MERIT Transform Performance on GPUs}

\begin{table}[t]
\centering
\caption{Speedup on GPUs compared to OpenCV, Parboil, and Caffe.}\label{tab:speed}
\begin{tabular}{llr}
\toprule
Kernels   & Note & Speedup\\
\midrule
Separable Filter & $k=3$  & 0.35\\
                 & $k=30$ & 1.42\\
\hline
Motion Estimation & & 6.51\\
\hline
Forward Propagation  & Ours $3+1s$ & 19.9\\
(kernel size+stride, & Ours $9+1s$ & 26.4\\
32 channels)         & Ours $3+2s$ & 1.80\\
                     & Ours $9+2s$ & 2.83\\
                     & cuDNN $3+1s$ & 100\\
                     & cuDNN $9+1s$ & 109\\
                     & cuDNN $3+2s$ & 27.1\\
                     & cuDNN $9+2s$ & 27.3\\
\bottomrule
\end{tabular}
\end{table}

\tabref{speed} shows the kernel performance gain by the MERIT transform.
For the separable filter in OpenCV, our approach is faster except for very small kernel sizes owing to the pre-computation overhead.
It is particularly interesting because, despite the simplistic nature of this kernel,
we are able to extract an extra 40\% performance gain over OpenCV.
Such extra gain comes from the address calculations \texttt{buffer[w*x+y]} in the code,
which takes several instructions while the actual MAC only takes one.
We replace all address calculation by constant memory lookup in all critical paths (\ie inner loops), reducing the instruction count to two.
While this technique is not only used by MERIT transform,
such optimization requires detailed knowledge about the GPUs,
which is tedious and obfuscates the program.
With a unified transform expressing various vision tasks,
these tasks can immediately benefit more from parallel computing with one optimized MERIT implementation.

On the other hand, although there remains a large gap between MERIT and the heavily-tuned cuDNN from NVIDIA, our implementation still surpasses Caffe, which is built on top of the heavily optimized cuBLAS linear algebra library, also from NVIDIA.
Also note that we hold an edge against Caffe for smaller strides and larger kernels, showing a performance pattern in line with cuDNN.

\subsection{MERIT-z Processor ASIC Implementation}

In \tabref{comp_all}, we compare MERIT-z with several dedicated DNN accelerators~\cite{eyeriss,maeri},
and \tabref{synthesize} shows the area and power breakdown of our processor.
We synthesize the $4$-TAU MERIT-z~($128$~ALUs) at 400~\si{MHz} to provide a similar computation power to these works.
Our architecture is similarly power efficient (per MAC) compared to these processors,
and it only uses half as much area.
While comparing our frontend numbers against published backend metrics plays in our favor,
it is clear that we compare favorably against these processors in terms of area and power. 
Since current MERIT-z implementation excludes the global buffer,
it has the highest area efficiency in this table.

These gains come from several factors.
First, MERIT transform allows us to use the much smaller single-port (SP) SRAMs, which are only half size compared to 2-port (TP) SRAMs, and this results in an overall area reduction of over 30\% with at most 3\% utilization rate drop.
Second, as discussed in \secref{proc:archcomp}, MERIT-z can achieve a high utilization rate with a comparatively smaller SRAM storage because of the efficient MERIT transform application over the butterfly network and omega network.
The simplicity of these classic circuit networks also allows us to run the chip at a much higher frequency because of their circuit routing efficiency.

To ensure the evaluation accuracy,
we adopt a cycle-accurate DRAM simulator called Ramulator~\cite{ramulator} and integrate it into our simulation through Verilog VPI.
We choose the 1-rank, 2-channel, DDR3~1600, 2Gbx8, and 3.2~GBps bandwidth in the simulation settings,
and the L1 cache size is set to 1~KB mainly to cover the misaligned DRAM access.
Also, we synthesize the MERIT-z using the TSMC 28~\si{nm} low-power RVT library with power simulation reported using Synopsys PrimeTime.

\begin{table}[t]
\centering
\caption{Comparison of DNN ASIC accelerators; we cite the backend metrics from \cite{maeri}.}\label{tab:comp_all}
\begin{tabular}{lrrrr}
\toprule
\footnotesize
& Systolic & Eyeriss & MAERI & MERIT-z\\
& (TPU) &&& (4-TAU)\\
\midrule
Process~(\si{nm})                 &  28 &  28 &  28 &   28\\
\#ALUs (MAC)                      & 168 & 168 & 168 &  128\\
\hline
Frequency~(\si{MHz})              & 200 & 200 &  200 & 400\\
Peak MAC per second               & 33G & 33G &  33G & 51G\\
Density~(ALU/\si{mm^2})           &  62 &  25 &   44 &  \textbf{86}\\
\hspace*{1em}Area~(\si{mm^2})     & 2.7 & 6.0 &  3.8 & 1.5\\
Power per MAC~(\si{pJ})           & \textbf{5.1} & 9.0 & 11.2 & 7.5\\
\hspace*{1em}Core Power~(\si{mW}) & 168 & 304 &  379 & 386\\
\bottomrule
\end{tabular}
\end{table}

\begin{table}[t]
\centering
    \caption{Area and power breakdown of a $4$-TAU MERIT-z.}\label{tab:synthesize}
\begin{tabular}{lrrr}
\toprule
Module & Logic (\%) & SRAM (\%) & Power (\%)\\
\midrule
Compute Pipeline & 15.7 & 22.0 & 57\\
Read Pipelines & 16.5 & 36.5 & 39\\
Write Pipeline + Misc. & 9.3 & 0.0 & 7\\
    &&&\\
Total (absolute) & \multicolumn{2}{c}{5.81~\si{mm^2} or} & 386~\si{mW}\\
& \multicolumn{2}{c}{4.08M gate count} & \\
\bottomrule
\end{tabular}
\end{table}

In \tabref{ex_nn}, we show the performance analysis of MERIT-z running AlexNet~\cite{alexnet} and VGG~\cite{vgg}.
In addition, we also benchmark a DNN-based saliency detection algorithm~\cite{saliency} consisting of additional DECONV layers and traditional image operations.
In \tabref{lot_nn}, we further benchmark several DNN layers developed in the state-of-the-art computer vision conferences~\cite{dilconv,pixel_shuf,flownet,mobilenet},
and these layers cannot be supported by any existing, single architecture because existing ones are designed for specific tasks.
All of these layers are expressed with MERIT transforms,
and the optimization techniques mentioned previously can be applied out-of-the-box.
For example, with the help of optimization techniques shown in \figref{prefetch},
we can achieve acceptable efficiency for depth-wise convolution even though this algorithm is mostly memory-bounded.

\begin{table}[t]
\centering
    \caption{DNN workloads on MERIT-z with 3.2~GBps DRAM.}\label{tab:ex_nn}
\begin{tabular}{lrr}
\toprule
Layer & Power~(\si{mW}) & Utilization\\
\midrule
\textbf{AlexNet (overall)}~\cite{alexnet} & 386 & 0.88\\
\hspace*{1em}CONV1+Pool2 & 334 & 0.88\\
\hspace*{1em}CONV2+Pool2 & 392 & 0.95\\
\hspace*{1em}CONV3       & 363 & 0.77\\
\hspace*{1em}CONV4       & 363 & 0.72\\
\hspace*{1em}CONV5       & 363 & 0.72\\
\hspace*{1em}FC1-3~(batch=32) & 360 & 0.92\\
    &&\\
\textbf{VGG-16D (overall)}~\cite{vgg} & 355 & 0.85\\
\hspace*{1em}CONV1            & 300 & 0.89\\
\hspace*{1em}CONV2-4+Pool2    & 363 & 0.95\\
\hspace*{1em}CONV5-13+Pool2   & 363 & 0.83\\
\hspace*{1em}FC1-3~(batch=32) & 360 & 0.92\\
    &&\\
\textbf{Saliency (overall)}~\cite{saliency} & 377 & 0.77\\
\hspace*{1em}CONV1+Pool3       & 412 & 0.72\\
\hspace*{1em}CONV2+Pool3       & 428 & 0.82\\
\hspace*{1em}CONV3+Pool3       & 420 & 0.65\\
\hspace*{1em}DECONV4           & 292 & 0.87\\
\hspace*{1em}DECONV5+threshold & 302 & 0.71\\
\hspace*{1em}Alpha blending    &  64 & 0.07\\
\bottomrule
\end{tabular}
\end{table}
\begin{table}[t]
\centering
\caption{MERIT-z efficiency for popular workloads.}\label{tab:lot_nn}
\begin{tabular}{lrrr}
\toprule
Computation & Output Tensor Size & Utilization\\
\midrule
Dilated~\cite{dilconv}                  & $64\times 64\times 32$           & 0.95\\
Pixel Shuffle~(ESPCN)~\cite{pixel_shuf} & $32\times 32\times 32$           & 0.96\\
Correlation~(FlowNet)~\cite{flownet}    & $64\times 64\times 11 \times 11$ & 0.74\\
Depthwise~(MobileNet)~\cite{mobilenet}  & $64\times 64\times 32$           & 0.63\\
GEMM (FC, RNN)                          & $256\times 128$                  & 0.92\\
Motion estimation                       & 720p, $8\times 8$ block          & 0.74\\
\bottomrule
\end{tabular}
\end{table}

\subsection{Scaling up MERIT-z Processor}\label{sec:eva:ablation}
\begin{figure*}[t]
\centering
\includegraphics[width=0.99\textwidth]{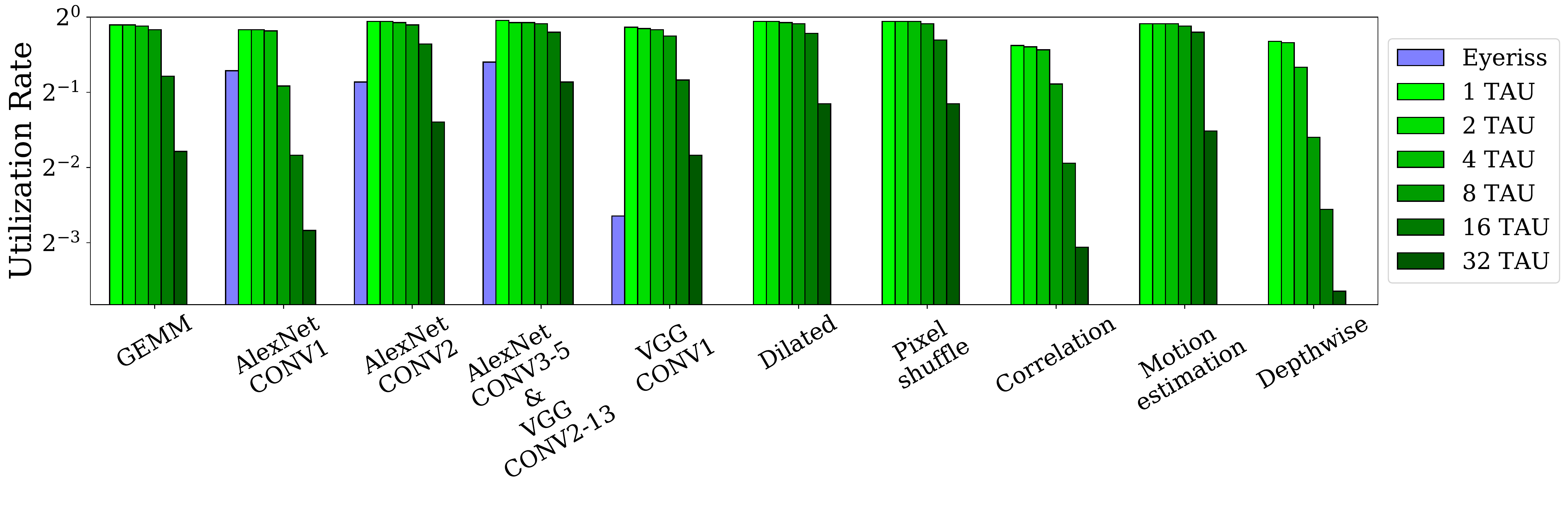}
\caption{\textbf{Utilization scaling of MERIT-z processor. } MERIT-z remains highly utilized among a variety of workloads up to $256$ ALUs, and for comparison we also calculate and plot the utilization rate of Eyeriss according to their processing latency.}\label{fig:comp_mc_speed}
\end{figure*}

The vector-architecture nature of MERIT-z means that the area can be scaled up fairly linearly when adding more TAUs to MERIT-z.
On the other hand, \figref{comp_mc_speed} shows the relative speedup with ALU counts between $32$ and $1024$, using the same DDR3 3.2~GBps simulation setting as before.
Most algorithms, except for the depth-wise convolution, scale up well up to $256$~ALUs, beyond which the applications become DRAM memory-bounded entirely.
Since MERIT-z only utilizes one level of the memory hierarchy,
we eliminate the need for moving data from the global prefetch buffer to local buffers,
which can be a bottleneck reported by \cite{eyeriss}.
Therefore, for DNN requiring multiple processing passes like VGG~CONV1,
we achieve particularly higher utilization than other works,
and such feature is essential for larger DNN workloads that are likely to appear in the future.

\section{Conclusion}\label{sec:con}
In this paper, we proposed the MERIT transform, a mathematical framework for transforming vision processing tasks into SIMD-friendly workloads by separating data movement from computations.
Since the core of algorithm optimization often lies in the data movement process, it allows us to write fast parallel kernels with very small lines of code.
Because this process is similar across different algorithms on the same processor, we created a library for CUDA GPUs so as to free programmers from the burden of repeated optimization efforts, such as thread tiling and shared memory access.
We also use these insights to design the MERIT-z processor.
The processor can perform a wide range of applications,
such as DNN, image processing, and traditional machine learning applications,
with comparable area and power efficiency to several dedicated DNN processors.
It uses classic circuit blocks such as butterfly networks to shuffle data to the ALUs efficiently.
Also, we have released both the CUDA library and MERIT-z processor implementations under the open-source GPL license.

The mathematical framework has proven to be a useful utility for verifying the efficiency of algorithms against processors.
We shall continue to identify more useful properties for the MERIT transform in order to exploit opportunities for data reuse between processors during execution.
We would also like to look into more matching patterns between transforms and circuits, provide more automatic means for generating these patterns,
such that it can become more powerful and useful for the parallel and scientific computing community.

\section*{Acknowledgment}
This work was partially sponsored by MediaTek Inc., Hsin-chu, Taiwan and MultiTek Inc., Hsin-chu, Taiwan.

\appendix[]
The CUDA implementation of MERIT can be found in Github under \verb|mediaic/UMI|, and the SystemVerilog implementation will soon be available under \verb|mediaic/MERIT|.

\bibliographystyle{IEEEtran}
\bibliography{bare_jrnl}

\end{document}